\documentclass[seceq]{ptptex}

\usepackage[dvips]{graphicx,color}
\usepackage{multirow}
\usepackage{bm}
\usepackage{wrapft}

\def\la{\mathrel{\mathpalette\fun <}}
\def\ga{\mathrel{\mathpalette\fun >}}
\def\fun#1#2{\lower3.6pt\vbox{\baselineskip0pt\lineskip.9pt
\ialign{$\mathsurround=0pt#1\hfil##\hfil$\crcr#2\crcr\sim\crcr}}}

\title{
Dynamical Relativistic Effects in Breakup Processes of Halo Nuclei
}

\author{
Kazuyuki \textsc{Ogata}$^1$ and
Carlos A. \textsc{Bertulani}$^2$
}

\inst{
$^1$Department of Physics, Kyushu University, Fukuoka 812-8581, Japan\\
$^2$Department of Physics, Texas A\&M University, Commerce, TX 75429, USA
}

\abst{The continuum-discretized coupled-channels (CDCC) method is used to study the breakup of weakly-bound nuclei at intermediate energies collisions. For large impact parameters, the Eikonal CDCC (E-CDCC) method was applied. The effects of
Lorentz contraction on the nuclear and Coulomb potentials have been investigated in details. Such effects tend to increase cross sections appreciably. We also show that, for loosely-bound nuclei, the contribution of the so-called close field is small and can be neglected.
}

\begin{document}
\maketitle

\section{Introduction}
\label{sec1}

The properties of unstable nuclei are one of the most important
subjects in science. The breakup reactions of such short-lived
nuclei provide us with plentiful information on their static and dynamical
features. Among them, the responses of unstable nuclei to electromagnetic fields have been intensively studied.
For this purpose, usually a heavy (highly charged)
target such as $^{208}$Pb is adopted
and many of the breakup experiments have been
performed at intermediate energies, say, 100--250 MeV per nucleon,
to minimize possible {\lq\lq}contaminations'' due to nuclear
breakup and higher-order breakup processes.

It is well known that a theoretical description of breakup reactions
at intermediate energies requires a relativistic treatment of the
reaction dynamics, though only a relativistic modification on the
kinematics has been usually included. The virtual photon theory, or,
equivalent photon method, is a fully relativistic reaction model
to describe the excitation of a projectile under electromagnetic fields
caused by a target nucleus. However, it cannot deal with the nuclear
breakup and higher-order breakup processes. To extract reliable
physics quantities such as B(E1)-values and asymptotic normalization
coefficients from breakup experiments, evaluation of the contributions
from the above mentioned processes is necessary.

In our recent Letter~\cite{OB09},
we developed a full coupled-channel calculation
including a relativistic treatment of not only the kinematics but
also the dynamics, based on the Continuum-Discretized Coupled-Channels
method (CDCC) \cite{CDCC} with the eikonal approximation, i.e., eikonal CDCC
(E-CDCC)~\cite{Og03,Og06}.
An essential ingredient of the relativistic CC calculation
was the proper treatment of nuclear and Coulomb coupling potentials between the projectile
and the target. We adopted the form of the Coulomb dipole and quadrupole
interactions shown in Ref.~\citen{Be05}, which was obtained from
a relativistic Li\'{e}nard-Wiechert potential with so-called
far-field approximation.
As for the nuclear potential, the conjecture of Feshbach and Zabek
\cite{FZ77} was adopted. We showed that the dynamical relativistic
corrections are responsible for an increase of about 15\% in the
breakup cross sections of $^8$B and $^{11}$Be projectiles by $^{208}$Pb at 250 MeV/nucleon.
Another important finding was that the nuclear breakup and higher-order
breakup processes were significant even at 250 MeV/nucleon.

In the present paper, we show a more detailed analysis of the dynamical
relativistic effects on the breakup cross sections of $^8$B and $^{11}$Be projectiles
by $^{208}$Pb at 250 MeV/nucleon.
First, we investigate the relativistic effects on the double differential
breakup cross section, which clarifies how relativity affects
the cross section at several emission angles $\theta$
and breakup energies $\epsilon$.
Differences in the relativistic effects between $^8$B and $^{11}$Be
breakup are also discussed.
Second, the contribution of the nuclear breakup and higher-order
processes are shown on the $\epsilon$-$\theta$ plane.
Third, a quantum mechanical correction in the breakup amplitude
obtained by E-CDCC is carried out, which enables one to perform
fully relativistic and quantum mechanical CC calculations of
breakup reactions at intermediate energies. Fourth,
the contribution from so-called close-fields, which are
neglected in obtaining the relativistic Coulomb interactions~\cite{Be05},
is evaluated by means of the first-order time-dependent theory.

The paper is constructed as follows. In \S\ref{sec2}, we summarize
some formulae of E-CDCC and show how to include the
dynamical relativistic corrections to the coupling potentials.
The contribution from the close-fields to the breakup process
is also formulated. In \S\ref{sec3}, we present our numerical results
and discuss the relevant physics. Finally, we give
a summary in \S\ref{sec4}.

\section{Formulation}
\label{sec2}

\subsection{Relativistic CDCC}
\label{sec2-1}

We start with the following nonrelativistic
E-CDCC equations \cite{Og03,Og06} for a three-body reaction
between a projectile P, consisting of
a core (C) and a valence nucleon v, and a target nucleus T:
\begin{equation}
\dfrac{i\hbar^2}{E_c}K_c^{(b)}(z)
\dfrac{d}{d z}\psi_{c}^{(b)}(z) = \sum_{c'}
{\mathfrak{F}}^{(b)}_{cc'}(z) \; {\cal R}^{(b)}_{cc'}(z) \;
\psi_{c'}^{(b)}(z) \ e^{i\left(K_{c'}-K_c \right) z}, \label{cceq4}
\end{equation}
where $c$ denotes the channel indices \{$i$, $\ell$, $m$\}; $i>0$
($i=0$) stands for the $i$th discretized-continuum (ground) state,
and $\ell$ and $m$\ are, respectively, the orbital angular momentum
between the constituents (C and v) of P and its
projection on the $z$-axis taken to be parallel to the incident
beam. We neglect the internal spins of C and v for
simplicity.
$b$ is the impact parameter (or
transverse coordinate) in the collision of P and T,
which is defined by $b=\sqrt{x^2+y^2}$ with ${\bm R} = (x,y,z)$,
the relative coordinate of P from T in the Cartesian representation.
Note that in Eq.~(\ref{cceq4})
$b$ is relegated to a superscript
since it is not a dynamical variable. The total energy and the
asymptotic wave number of P are denoted by $E_c$ and $K_c$,
respectively, and ${\cal R}_{cc'}^{(b)}(z)=(K_{c'} R-K_{c'}
z)^{i\eta_{c'}}/ (K_c R-K_c z)^{i\eta_c}$ with $\eta_c$ the
Sommerfeld parameter. The local wave number $K_c^{(b)}(z)$ of P is
defined by energy conservation as
\begin{equation}
E_c
=
\sqrt{(m_{\rm P}c^2)^2+\left[\hbar c K_c^{(b)}(z)\right]^2}
+
\dfrac{Z_{\rm P}Z_{\rm T}e^2}{R},
\label{klcl}
\end{equation}
where $m_{\rm P}$ is the mass of P and $Z_{\rm P}e$ ($Z_{\rm T}e$)
is the charge of P (T).
The reduced coupling potential ${\mathfrak{F}}^{(b)}_{cc'}(z)$ is given by
\begin{equation}
{\mathfrak{F}}^{(b)}_{cc'}(z)
=
{\cal F}^{(b)}_{cc'}(z)
-\dfrac{Z_{\rm P}Z_{\rm T}e^2}{R}\delta_{cc'},
\label{FF1}
\end{equation}
where
\begin{equation}
{\cal F}^{(b)}_{cc'}(z)
=
\left\langle
\Phi_{c}
|
U_{\rm CT}+U_{\rm vT}
|
\Phi_{c'}
\right\rangle_{\bm \xi}
=
{\cal F}^{{\rm nucl}(b)}_{cc'}(z)+{\cal F}^{{\rm Coul}(b)}_{cc'}(z),
\label{FF2}
\end{equation}
\begin{eqnarray}
{\cal F}^{{\rm nucl}(b)}_{cc'}(z)
&=&
\left\langle
\Phi_{c}
|
U^{\rm nucl}_{\rm CT}+U^{\rm nucl}_{\rm vT}
|
\Phi_{c'}
\right\rangle_{\bm \xi},
\\
{\cal F}^{{\rm Coul}(b)}_{cc'}(z)
&=&
\left\langle
\Phi_{c}
|
U^{\rm Coul}_{\rm CT}+U^{\rm Coul}_{\rm vT}
|
\Phi_{c'}
\right\rangle_{\bm \xi}.
\label{NRC}
\end{eqnarray}
$\Phi_c({\bm \xi})$ denotes the internal wave functions of P,
with ${\bm \xi}$ the coordinate of v relative to C,
and $U_{\rm CT}$ ($U_{\rm vT}$) is the
potential between C (v) and T consisting of nuclear
and Coulomb parts.
Furthermore, in actual calculations, we use the multipole expansion
for each term on the right-hand-side of Eq.~(\ref{FF2}):
\begin{eqnarray}
{\cal F}^{{\rm nucl}(b)}_{cc'}(z)&=&
\sum_\lambda {\cal F}^{{\rm nucl}(b)}_{cc',\lambda}(z),
\\
{\cal F}^{{\rm Coul}(b)}_{cc'}(z)&=&
\sum_\lambda {\cal F}^{{\rm Coul}(b)}_{cc',\lambda}(z).
\end{eqnarray}
The explicit form of the multipoles is given in Ref.~\citen{Og06}.

In Ref.~\citen{Be05}, the relativistic form of
the electric dipole (E1) and quadrupole (E2) interactions are
given by
\begin{equation}
V^{\rm rel}_{{\rm E1}\mu}(b,z,\mathbf{\hat{\mbox{\boldmath$\xi$}}})
=\sqrt{\frac{2\pi}{3}}\xi Y_{1\mu}\left(  \mathbf{\hat
{\mbox{\boldmath$\xi$}}}\right)  \frac{\gamma Z_{\rm T}ee_{\rm E1}}{\left(
b^{2}+\gamma ^{2}z^{2}\right)^{3/2}}\left\{
\begin{array}
[c]{c}%
\mp b,\ \ (\mathrm{if}\ \ \ \mu=\pm1)\\
\sqrt{2}z\ \ (\mathrm{if}\ \ \ \mu=0)\
\end{array}
\right.,  \label{relE1}%
\end{equation}
\begin{align}
V^{\rm rel}_{{\rm E2}\mu}(b,z,\mathbf{\hat{\mbox{\boldmath$\xi$}}})
= &  \sqrt{\frac{3\pi}{10}}\xi^{2}Y_{2\mu}\left(
\mathbf{\hat {\mbox{\boldmath$\xi$}}}\right)  \frac{\gamma
Z_{\rm T}ee_{\rm E2}}{\left(  b^{2}+\gamma
^{2}z^{2}\right)  ^{5/2}}\nonumber\\
&  \times\left\{
\begin{array}
[c]{c}%
b^{2},\ \ \ \ (\mathrm{if}\ \ \ \mu=\pm2)\\
\mp(\gamma^2+1)bz,\ \ \ \ (\mathrm{if}\ \ \ \mu=\pm1)\\
\sqrt{2/3}\left(  2\gamma^{2}z^{2}-b^{2}\right)  \ \ \ \ (\mathrm{if}%
\ \ \ \mu=0)\
\end{array}
\right.,  \label{relE2}%
\end{align}
where $e_{{\rm E}\lambda}=[Z_{\rm v}(A_{\rm
C}/A_{\rm P})^\lambda +Z_{\rm C}(-A_{\rm v}/A_{\rm P})^\lambda]e$
are effective charges for $\lambda=1$ and 2 multipolarities for the
breakup of ${\rm P}\rightarrow{\rm C}+{\rm v}$;
$A_j$ ($j={\rm C}$, v, P) represents the mass number of the particle $j$.
The Lorentz
contraction factor is denoted by $\gamma=\left(
1-v^{2}/c^{2}\right)^{-1/2}$, where $v$ is the velocity of P.
Equations (\ref{relE1}) and (\ref{relE2}) are obtained with
so-called far-field approximation \cite{EB02}, i.e., $R$ is assumed
to be always larger than both $\xi A_{\rm v}/A_{\rm P}$
and $\xi A_{\rm C}/A_{\rm P}$.
Derivation of these equations
is described in detail in Ref.~\citen{BCG03}.
In addition, for magnetic dipole excitations (not considered here),
\begin{equation}
V^{\rm rel}_{{\rm M1}\mu}(b,z,\mathbf{{\mbox{\boldmath$\xi$}}})
=i\sqrt{\frac{2\pi}{3}}
\bar{M}_{1\mu}\left(  \mathbf{{\mbox{\boldmath$\xi$}}}\right)
{v \over c}
\frac{\gamma Z_{\rm T}e}{\left(
b^{2}+\gamma ^{2}z^{2}\right)^{3/2}}\left\{
\begin{array}
[c]{c}%
\pm b,\ \ (\mathrm{if}\ \ \ \mu=\pm1)\\
0\ \ (\mathrm{if}\ \ \ \mu=0)\
\end{array}
\right.,  \label{relM1}%
\end{equation}
where
$\bar{M}_{1\mu}(  \mathbf{{\mbox{\boldmath$\xi$}}})$
is the intrinsic M1 operator.

One may find easily that $V^{\rm rel}_{{\rm E1}\mu}$
and $V^{\rm rel}_{{\rm E2}\mu}$ are, respectively, obtained
from their nonrelativistic expressions
$V_{{\rm E1}\mu}$ and $V_{{\rm E2}\mu}$ as
\[
V^{\rm rel}_{{\rm E1}\pm1}(b,z,\mathbf{\hat{\mbox{\boldmath$\xi$}}})
=
\gamma
V_{{\rm E1}\pm1}(b,\gamma z,\mathbf{\hat{\mbox{\boldmath$\xi$}}}),
\quad
V^{\rm rel}_{{\rm E1}0}(b,z,\mathbf{\hat{\mbox{\boldmath$\xi$}}})
=
V_{{\rm E1}0}(b,\gamma z,\mathbf{\hat{\mbox{\boldmath$\xi$}}}),
\]
\[
V^{\rm rel}_{{\rm E2}\pm2}(b,z,\mathbf{\hat{\mbox{\boldmath$\xi$}}})
=
\gamma
V_{{\rm E2}\pm2}(b,\gamma z,\mathbf{\hat{\mbox{\boldmath$\xi$}}}),
\quad
V^{\rm rel}_{{\rm E2}\pm1}(b,z,\mathbf{\hat{\mbox{\boldmath$\xi$}}})
=
\frac{\gamma^2+1}{2}
V_{{\rm E2}\pm1}(b,\gamma z,\mathbf{\hat{\mbox{\boldmath$\xi$}}}),
\]
\[
V^{\rm rel}_{{\rm E2}0}(b,z,\mathbf{\hat{\mbox{\boldmath$\xi$}}})
=
\gamma
V_{{\rm E2}0}(b,\gamma z,\mathbf{\hat{\mbox{\boldmath$\xi$}}}).
\]
Thus, we can include the dynamical relativistic corrections in E-CDCC
by carrying out the replacement
\begin{equation}
{\cal F}^{{\rm Coul}(b)}_{cc',\lambda}(z)
\rightarrow \gamma f_{\lambda,m-m'}
{\cal F}^{{\rm Coul}(b)}_{cc',\lambda}(\gamma z)
\label{FF4}
\end{equation}
with
\begin{equation}
f_{\lambda,\mu}
=
\left\{
\begin{array}{cl}
1/\gamma, & \quad (\lambda=1, \mu=0) \\
(\gamma^2+1)/(2\gamma),   & \quad (\lambda=2, \mu=\pm1) \\
1        & \quad ({\rm otherwise}) \\
\end{array}
\right..
\label{FF5}
\end{equation}
Correspondingly, we use
\begin{equation}
\dfrac{Z_{\rm P}Z_{\rm T}e^2}{R}\delta_{cc'}
\rightarrow
\gamma\dfrac{Z_{\rm P}Z_{\rm T}e^2}{\sqrt{b^2+(\gamma z)^2}}\delta_{cc'}
\label{FF6}
\end{equation}
in Eqs.~(\ref{klcl}) and (\ref{FF1}).
For the nuclear coupling potential
${\cal F}^{{\rm nucl}(b)}_{cc',\lambda}(z)$, we adopt
\begin{equation}
{\cal F}^{{\rm nucl}(b)}_{cc',\lambda}(z)
\rightarrow \gamma
{\cal F}^{{\rm nucl}(b)}_{cc',\lambda}(\gamma z)
\label{FF7}
\end{equation}
following the conjecture of Feshbach and Zabek \cite{FZ77}.
In the present CC calculation, the Lorentz contraction factor
$\gamma$ may have channel dependence, i.e., $\gamma=E_c/(m_{\rm
P}c^2)$, which we approximate using the value in the incident
channel, i.e., $E_0/(m_{\rm P}c^2)$. This can be justified
since the energy transfer to the projectile is significantly
small compared with the incident energy, in the reactions
considered here.

Solving Eq.~(\ref{cceq4}) under the boundary condition
\begin{equation}
\lim_{z \to -\infty}\psi_{c}^{(b)}(z)=\delta_{c0},
\end{equation}
where $0$ denotes the incident channel, one obtains the
following form of the eikonal scattering amplitude:
\begin{equation}
f^{\rm E}_{c0}
=
f_{c0}^{\rm Ruth}\delta_{c0}
+
\dfrac{2\pi}{i K_0}
\sum_L
f'^{\rm E}_{L;c0}\;
Y_{L \,m-m_0}(\hat{\bf K}'_c),
\label{fC7}
\end{equation}
where $f_{c0}^{\rm Ruth}$ is the Rutherford amplitude. The
partial scattering amplitude $f'^{\rm E}_{L;c0}$ is defined by
\begin{equation}
f'^{\rm E}_{L;c0}
=
\dfrac{K_0}{K_{c}}
{\cal H}^{(b_{c;L})}_c
\sqrt{\dfrac{2L+1}{4\pi}}
i^{(m-m_0)}
\left[
{\cal S}_{c0,L}
-
\delta_{c0}
\right],
\label{fC8}
\end{equation}
where
\begin{equation}
{\cal S}_{c0,L} \equiv
\displaystyle
{\lim_{z \to \infty}}
\psi_{c}^{(b_{c;L})}(z)
\end{equation}
with $b_{c;L} = (L + 1/2)/K_c$, and
${\cal H}_{c,L} \equiv \exp[2i\eta_c\ln{(L+1/2)}]$.

The quantum mechanical (QM) correction in the scattering amplitude
can be performed if one replaces $f'^{\rm E}_{L;c0}$ for
small $L$, say, $L < L_{\rm C}$, by the QM partial amplitude
obtained with the conventional QM CDCC:
\begin{eqnarray}
f'^{\rm Q}_{L;c0}
&\equiv&
\sum_{J=|L-\ell|}^{L+\ell}
\sum_{L_0=|J-\ell_0|}^{J+\ell_0}
\sqrt{\dfrac{2L_0+1}{4\pi}}
(\ell_0 m_0 L_0 0 | J m_0)
(\ell m L \;m_0\!\!-\!m| J m_0)
\nonumber \\
& &
\times
(S_{i L \ell,i_0 L_0 \ell_0}^{J}-\delta_{i i_0}
\delta_{L L_0}\delta_{\ell \ell_0})
e^{i(\sigma_L+\sigma_{L_0})}
(-)^{m-m_0},
\label{fCq}
\end{eqnarray}
where $\sigma_L$ is the Coulomb phase shift
and $J$ is the total angular momentum of the three-body system.
This correction is valid if $f'^{\rm E}_{L;c0}=f'^{\rm Q}_{L;c0}$
for $L\ge L_{\rm C}$; this is in fact the definition of $L_{\rm C}$.
Note that in a full QM calculation, i.e., without
the eikonal approximation, inclusion of the dynamical relativistic
corrections in the coupling potentials is very complicated
and actually inconsistent with the formalism. Fortunately, however,
it was shown in our previous Letter \cite{OB09} that the dynamical
relativistic corrections are necessary only for large $L$,
where the scattering processes
are well described by the eikonal approximation. Thus, using the
following scattering amplitude
\begin{equation}
f_{c0}
=
f_{c0}^{\rm Ruth}\delta_{c0}
+
\dfrac{2\pi}{i K_0}
\sum_{L < L_{\rm C}}
f'^{\rm Q}_{L;c0}\;
Y_{L \,m-m_0}(\hat{\bf K}'_c)
+
\dfrac{2\pi}{i K_0}
\sum_{L \ge L_{\rm C}}
f'^{\rm E}_{L;c0}\;
Y_{L \,m-m_0}(\hat{\bf K}'_c),
\label{fh}
\end{equation}
with the dynamical relativistic correction in the evaluation
of $f'^{\rm E}_{L;c0}$, one can carry out a relativistic QM
calculation including both nuclear and Coulomb couplings,
and also all higher-order processes, i.e., a {\it relativistic
CDCC} calculation.

\subsection{Contribution from close-fields to Coulomb breakup processes}
\label{sec2-2}

As mentioned in \S\ref{sec2-1}, we have used the relation between the
relativistic and nonrelativistic E$\lambda$ ($\lambda=1$ or 2)
interactions that is appropriate for the far-field collisions.
To verify this assumption, we evaluate the contribution from
the E1 close-field collisions in the following, on the basis of
the first-order time-dependent theory.

We start with the multipole-expansion form of the Li\'{e}nard-Wiechert
potential for close-field collisions given by Eq.~(9b) of Ref.~\citen{EB02}:
\begin{equation}
\phi^{\mathrm{close}}(\mathbf{r},\mathbf{R})
=\sum_{\lambda \mu \mu^{\prime}}
4\pi Y_{\lambda\mu}^{\ast}\left(
\mathbf{\hat{r}}\right)
  \sum_{\Lambda=0,2,4...}i^{\Lambda}R_{\lambda,\lambda-\Lambda
}\left(  r,R\right)
 A_{\lambda\mu,\lambda-\Lambda \mu^{\prime}}\left(  \beta\right)
Y_{\lambda-\Lambda,\mu^{\prime}}\left(  \mathbf{\hat{R}}\right)  ,
\label{LWpot}
\end{equation}
where
$\mathbf{r}$ is the coordinate of a point
charge in P relative to the center of mass (c.m.)
of P, and $\mathbf{R}$ is the relative coordinate between
the c.m. of P and another point charge in T.
In Eq.~\eqref{LWpot}, $R_{\lambda,\lambda^{\prime}}$ and
$A_{\lambda\mu,\lambda^{\prime}\mu^{\prime}}$ are given by
\begin{equation}
R_{\lambda,\lambda^{\prime}}\left(  r,R\right)
=
\frac
{1}{\sqrt{rR}}\int_{0}^{\infty}\frac{dq}{q}J_{\lambda+1/2}\left(
qr\right)  J_{\lambda^{\prime}+1/2}\left(  qR\right)  ,
\end{equation}%
\begin{equation}
A_{\lambda\mu,\lambda^{\prime}\mu^{\prime}}\left(  \beta\right)
=
\sum_{X \mathrm{\ even}}g_{X
}\left(  \beta\right)
 \left\langle Y_{\lambda^{\prime}\mu^{\prime}}\left(
\mathbf{\hat{q}}\right)  \left\vert
P_{X}\left(  \theta_{q}\right)
\right\vert Y_{\lambda\mu}\left(  \mathbf{\hat{q}}\right)  \right\rangle,
\end{equation}
where
\begin{equation}
g_{X}\left(  \beta\right)  =\frac{\left(  2X+1\right)  }{\beta
}Q_{X}\left(  \beta^{-1}\right)  ,\quad Q_{X}\left(  z\right)
=\frac{1}{2}\int_{-1}^{1}dt\frac{P_{X}\left(  t\right)  }{z-t}
\end{equation}
with $\beta=v/c$ and $\mathbf{q}$ is the momentum conjugate to the relative
coordinate $\mathbf{r}-\mathbf{R}$.

If we consider E1 transitions, i.e., $\lambda=1$,
we see that $\Lambda=0$ is only allowed, since $\Lambda$ must be even
and $\lambda-\Lambda$ is not negative. We then have%
\begin{equation}
\phi^{\mathrm{close}}_{\left(  \lambda=1\right)  }(\mathbf{r},\mathbf{R})
=\sum_{\mu\mu^{\prime}}4\pi
Y_{1\mu}^{\ast}\left(  \mathbf{\hat{r}}\right)  R_{1,1}
\left(  r,R\right)  A_{1\mu,1\mu^{\prime}}\left(  \beta\right)
Y_{1\mu^{\prime}}\left(
\mathbf{\hat{R}}\right)
\end{equation}
with%
\begin{equation}
A_{1\mu,1\mu^{\prime}}\left(  \beta\right)
=
\left\{  g_{0}\left(  \beta\right)  -\sqrt{\frac{2}{5}}\left(
1\mu 20|1\mu\right)  g_{2}\left(  \beta\right)  \right\}
\delta_{\mu\mu^{\prime}}.
\end{equation}
The diagonal radial part $R_{\lambda,\lambda}
\left(  r,R\right)  $\ is given by
Eq. (6) of Ref.~\citen{EB02} and we have%
\begin{equation}
R_{1,1}\left(  r,R\right)  =\frac{1}{3}\frac{R}{r^{2}}.
\end{equation}
Note that we assume here $r>R$.
From Eqs.~(4b) and (4c) of Ref.~\citen{EB02}, one can obtain explicit form
of $g_{0}\left(  \beta\right)$ and $g_{2}\left(  \beta\right)$:
\begin{equation}
g_{0}\left(  \beta\right)
=
\frac{1}{\beta}
\frac{1}{2\beta}\ln\left(  \frac{1+\beta}{1-\beta}\right),
\end{equation}
\begin{equation}
g_{2}\left(  \beta\right)
=
\frac{5}{4\beta}\left(\frac{3}{\beta^2}-1\right)  \ln\left(
\frac{1+\beta}{1-\beta}\right)  -\frac{15}{2\beta^2}.
\end{equation}
Thus, the relativistic Coulomb dipole interaction
between T and a proton inside P for close-field collisions
is given by
\begin{equation}
\bar{V}^{\mathrm{close}}_{1\mu}\left(  \mathbf{r},\mathbf{R}\right)
=\sqrt{\frac{2\pi}{3}}Z_{\rm T}e^{2}\frac{1}{r^{2}}Y_{1\mu}^{\ast}
\left(
\mathbf{\hat{r}}\right)  \left\{  g_{0}\left(  \beta\right)  +c_{\mu}%
g_{2}\left(  \beta\right)  \right\}  \left\{
\begin{array}
[c]{cc}%
\sqrt{2}z & \quad\mathrm{if\quad}\mu=0\\
\mp b & \quad\mathrm{if\quad}\mu=\pm1
\end{array}
\right.,
\label{VbarClose}
\end{equation}
where%
\begin{equation}
c_{0}=\frac{2}{5},\quad c_{\pm1}=-\frac{1}{5}.
\end{equation}
For far-field E1 collisions, we use the expression
$\bar{V}^{\mathrm{far}}_{1\mu}\left(  \mathbf{r},\mathbf{R}\right)=
V^{{\rm rel}*}_{{\rm E1}\mu}(b,z,\mathbf{\hat{\mbox{\boldmath$\xi$}}
=\hat{\bf r}})$
in Eq.~\eqref{relE1} with $e_{\rm E1}$ replaced by $e$.

We estimate the importance of close- and far-field collisions
by using first-order
time-dependent theory; we consider the contribution of one proton in P to
the breakup process.
The transition amplitudes due to the E1 field, from the initial
state $\Phi_{0}(\mbox{\boldmath$\xi$})$ to the final one
$\Phi_{c_f}(\mbox{\boldmath$\xi$})$, are given by%
\begin{equation}
a_{1\mu}\left(  b\right)
=
a_{1\mu}^{\mathrm{close}}\left(  b\right)
+
a_{1\mu}^{\mathrm{far}}\left(  b\right),
\nonumber
\end{equation}
\begin{equation}
a_{1\mu}^{\mathrm{close}}\left(  b\right)
=
\frac{1}{i\hbar}\int_{-\infty
}^{\infty}e^{i\omega t}\left
\langle \Phi_{c_f}\left(  \mbox{\boldmath$\xi$}\right)
\left\vert \bar{V}^{\mathrm{close}}_{1\mu}\left(  \mathbf{r},\mathbf{R}\right)  \right\vert \Phi_{0}\left(  \mbox{\boldmath$\xi$}\right)  \right\rangle
_{r>R}dt,
\end{equation}%
\begin{equation}
a_{1\mu}^{\mathrm{far}}\left(  b\right)
=
\frac{1}{i\hbar}\int_{-\infty
}^{\infty}e^{i\omega t}\left\langle \Phi_{c_f}\left(  \mbox{\boldmath$\xi$}\right)
\left\vert \bar{V}^{\mathrm{far}}_{1\mu}\left(  \mathbf{r},\mathbf{R}\right)
\right\vert \Phi_{0}\left(  \mbox{\boldmath$\xi$}\right)  \right\rangle _{r<R%
}dt,
\end{equation}
where $\omega$ represents the energy transfer as
$\hbar\omega\equiv E_{c_f}-E_{0}$.
Since we assume that the projectile moves along the $z$-axis with
a constant velocity $v$,
we immediately find $t=z/v$ and%
\begin{equation}
\omega t=\frac{E_{c_f}-E_{0}}{\hbar c}%
\frac{z}{\beta}.
\label{omegat}
\end{equation}

Using Eqs.~(\ref{VbarClose})--(\ref{omegat}), we obtain the explicit form of
$a_{1\mu}^{\mathrm{close}}$\ and $a_{1\mu}^{\mathrm{far}}$:
\begin{eqnarray}
a_{1\mu}^{\mathrm{close}}\left(  b\right)
&=&
\frac{1}{i\hbar v}\sqrt{\frac{2\pi
}{3}}Z_{\rm T}e^{2}\left\{  g_{0}\left(  \beta\right)  +c_{\mu}g_{2}\left(
\beta\right)  \right\}
\nonumber \\
& &
\times
\int_{-\infty}^{\infty}dz\exp\left(  i\frac
{E_{c_f}-E_{0}}{\hbar c}\frac{z}{\beta}\right)  \mathcal{M}_{1}^{\mathrm{close}%
}\left(  R\right)  \left\{
\begin{array}
[c]{cc}%
\sqrt{2}z & \quad\mathrm{if\quad}\mu=0\\
\mp b & \quad\mathrm{if\quad}\mu=\pm1
\end{array}
\right.  ,
\nonumber
\end{eqnarray}
\begin{eqnarray}
a_{1\mu}^{\mathrm{far}}\left(  b\right)
&=&
\frac{1}{i\hbar v}\sqrt{\frac{2\pi}%
{3}}\gamma Z_{\rm T}e^{2}
\nonumber \\
&&\times
\int_{-\infty}^{\infty}dz\exp\left(  i\frac{E_{c_f}-E_{0}%
}{\hbar c}\frac{z}{\beta}\right)  \frac{\mathcal{M}_{1}^{\mathrm{far}}\left(
R\right)  }{\left(  b^{2}+\gamma^{2}z^{2}\right)  ^{3/2}}\left\{
\begin{array}
[c]{cc}%
\sqrt{2}z & \quad\mathrm{if\quad}\mu=0\\
\mp b & \quad\mathrm{if\quad}\mu=\pm1
\end{array}
\right. ,
\nonumber
\end{eqnarray}
where
\begin{equation}
\mathcal{M}_{1\mu}^{\mathrm{close}}\left(  R\right)
\equiv
\left\langle \Phi_{c_f}\left(  \mbox{\boldmath$\xi$}\right)
\left\vert \frac{1}{r^{2}}Y_{1\mu}^{\ast}\left(  \mathbf{\hat{r}}\right)
\right\vert \Phi_{0}\left(\mbox{\boldmath$\xi$}\right)
\right\rangle _{r>R},
\end{equation}
\begin{equation}
\mathcal{M}_{1\mu}^{\mathrm{far}}\left(  R\right)
\equiv
\left\langle \Phi_{c_f}\left(  \mbox{\boldmath$\xi$}\right)
\left\vert rY_{1\mu}^{\ast
}\left(  \mathbf{\hat{r}}\right)  \right\vert
\Phi_{0}\left( \mbox{\boldmath$\xi$} \right)  \right\rangle _{r<R}.
\end{equation}
By comparing $a_{1\mu}\left(  b\right)$ with
$a_{1\mu}^{\mathrm{far}}\left(  b\right)$, we can evaluate how large
far-field collisions contribute to the total breakup amplitude.

As a typical reaction, let us consider the Coulomb breakup
 of $^8$B. The ground state wave function reads
\begin{equation}
\Phi_{0}(\mbox{\boldmath$\xi$})
=u_{0}(\xi)
Y_{1 m_0}(\hat{\mbox{\boldmath$\xi$}}).
\label{phi0}
\end{equation}
For the final state of $^8$B after the breakup we choose a
discretized continuum state with $l=0$, i.e.,
\begin{equation}
\Phi_{c_f}(\mbox{\boldmath$\xi$})
=u_{c_f}(\xi)
Y_{00}(\hat{\mbox{\boldmath$\xi$}}).
\label{phif}
\end{equation}
Here, $u_{c}(\xi)$ ($c=0$ or $c_f$) is the radial part
of $\Phi_{c}(\mbox{\boldmath$\xi$})$.
Using Eqs.~(\ref{phi0}) and (\ref{phif}) together with
$\mathbf{r}=(7/8)\mbox{\boldmath$\xi$}$, we have
\begin{equation}
\mathcal{M}_{1\mu}^{\mathrm{close}}\left(  R\right)
=\frac{1}{\sqrt{4\pi}}\left(  \frac{8}{7}\right)^{2}
\delta_{\mu,m_0}\int_{8R/7}^{\infty}u_{c_f}
\left(  \xi\right)  u_{0}\left(\xi\right)  d\xi,
\end{equation}
\begin{equation}
\mathcal{M}_{1\mu}^{\mathrm{far}}\left(  R\right)
=\frac{1}{\sqrt{4\pi}}\frac{7}{8}
\delta_{\mu,m_0}\int_{0}^{8R/7}u_{c_f}\left(
\xi\right)  u_{0}\left(  \xi\right)  \xi^{3}d\xi.
\end{equation}

\section{Results and discussions}
\label{sec3}

\subsection{Numerical input}
\label{sec3-1}

We use the same set of inputs as in Ref.~\citen{OB09} for the CDCC
calculation of $^{8}$B and $^{11}$Be breakup reactions with
$^{208}$Pb target at 250 MeV/nucleon.
The internal Hamiltonian of $^{8}$B and $^{11}$Be
are the same as in Ref.~\citen{Hussein}
except that we neglect the spin of the proton and
therefore we change accordingly the depth of the $p$-$^7$Be potential to reproduce the
proton separation energy of 137 keV. Note that in Ref.~\citen{Hussein},
and also in this work, the spins of neutron and C are disregarded.
We account for s, p, d, and f-states
in $p$-$^{7}$Be and $n$-$^{11}$Be.
The maximum value of the relative wave number, $k_{\rm max}$,
is 0.66 fm$^{-1}$ for all such states.
The number of discretized continuum states
is 20 and 10 for the s-waves and the other waves, respectively, for
both $^{8}$B and $^{11}$Be.
The optical potentials for
the constituents of P, i.e., $p$, $n$, $^7$Be, and $^{11}$Be, on
$^{208}$Pb are
the same as in Table I of Ref.~\citen{Hussein}.
The maximum
value used for the internal coordinate $\xi$ is 200 fm, and the
maximum impact parameter is 400 fm for both
$^8$B and $^{11}$Be breakup reactions.

%
\begin{figure}[b]
\centerline{
\includegraphics[width=70mm,keepaspectratio]{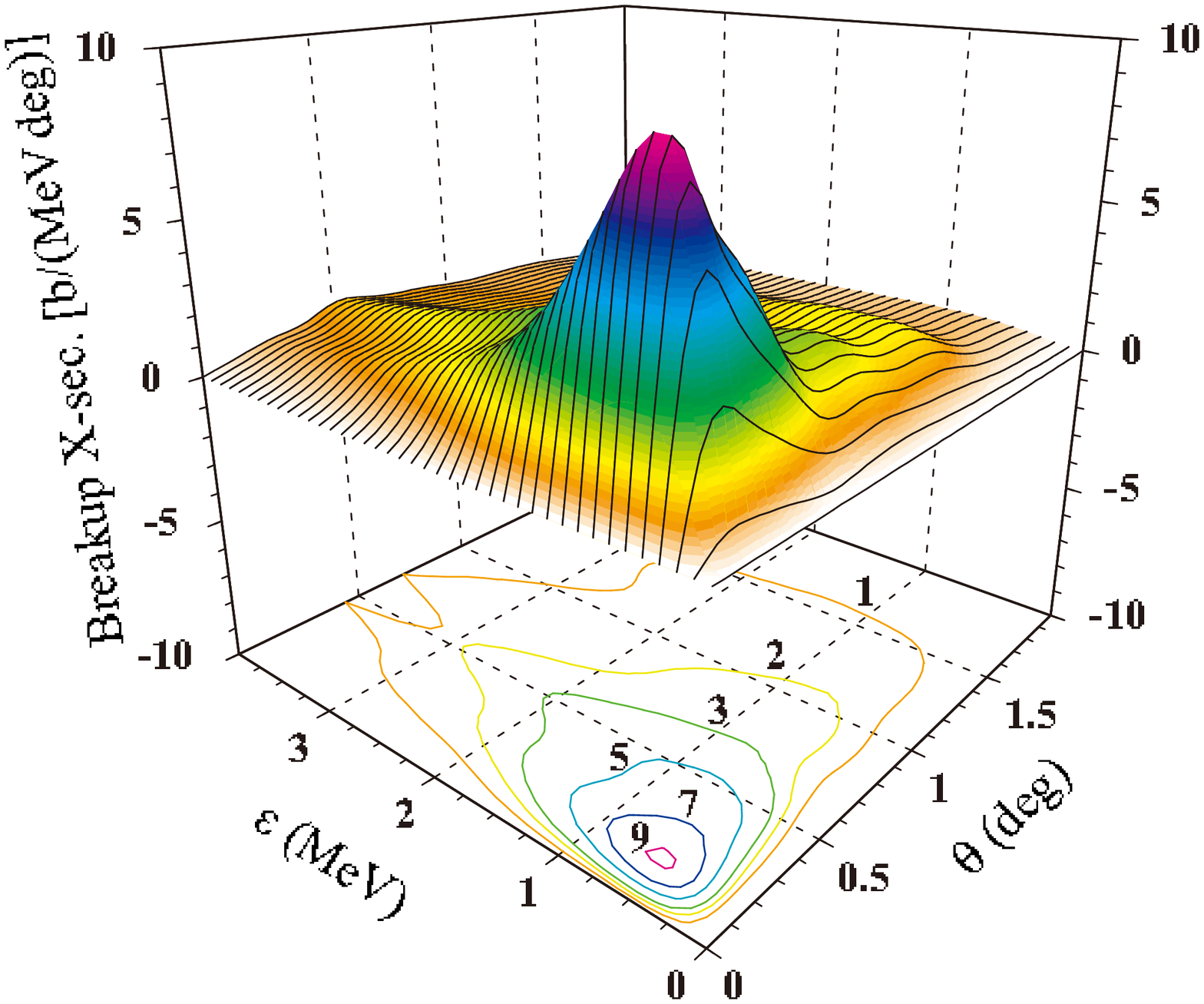}
\includegraphics[width=70mm,keepaspectratio]{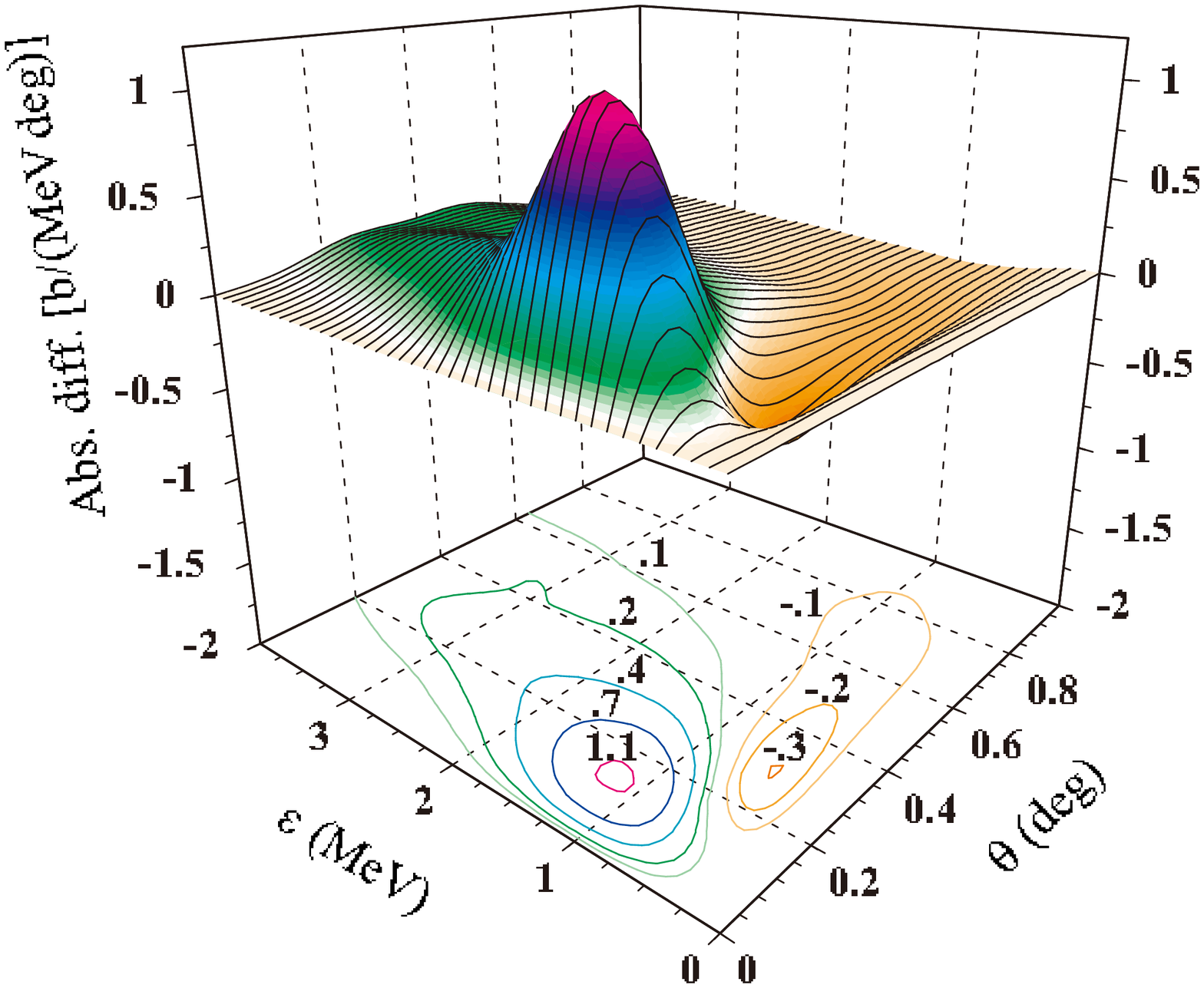}
} \caption{\label{fig1} Double differential breakup cross section
(DDBUX) for $^8$B$+^{208}$Pb at 250 MeV/nucleon including dynamical
relativistic corrections (left panel), and its difference from the
calculation without relativistic corrections (right panel). }
\end{figure}
Note that in the present
work, a {\it relativistic} calculation means a calculation which
includes dynamical relativistic corrections in the nuclear and
Coulomb coupling potentials. Also, a relativistic treatment of the
kinematics is adopted in all the calculations shown below.
Furthermore, we add quantum mechanical (QM) corrections in the
breakup amplitudes obtained with the relativistic E-CDCC as described
in \S\ref{sec2-1}; we use the results of fully QM but nonrelativistic
CDCC for the amplitudes corresponding to $b \le 18$ fm for both
reactions. This prescription has been numerically tested
and the results are shown
in \S\ref{sec3-4}. We assume, as in Ref.~\citen{OB09}, the
far-field approximation to obtain the relativistic
form of the Coulomb interaction between
each constituent of the projectile and the target nucleus. The
validity of this assumption is evaluated in \S\ref{sec3-5}.

\subsection{Dynamical relativistic effects on breakup cross section}
\label{sec3-2}

%
\begin{figure}[t]
\centerline{
\includegraphics[width=70mm,keepaspectratio]{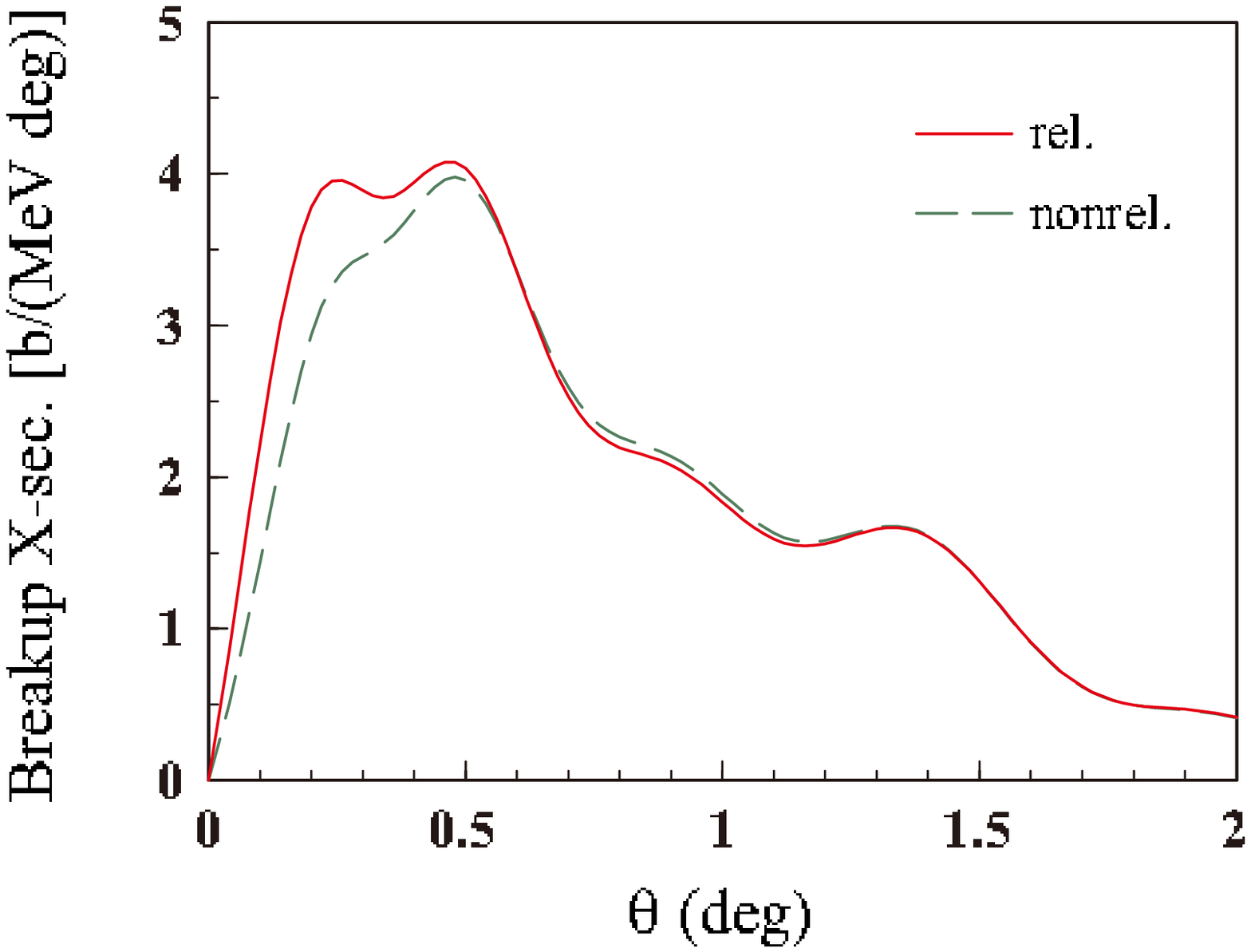}
\includegraphics[width=70mm,keepaspectratio]{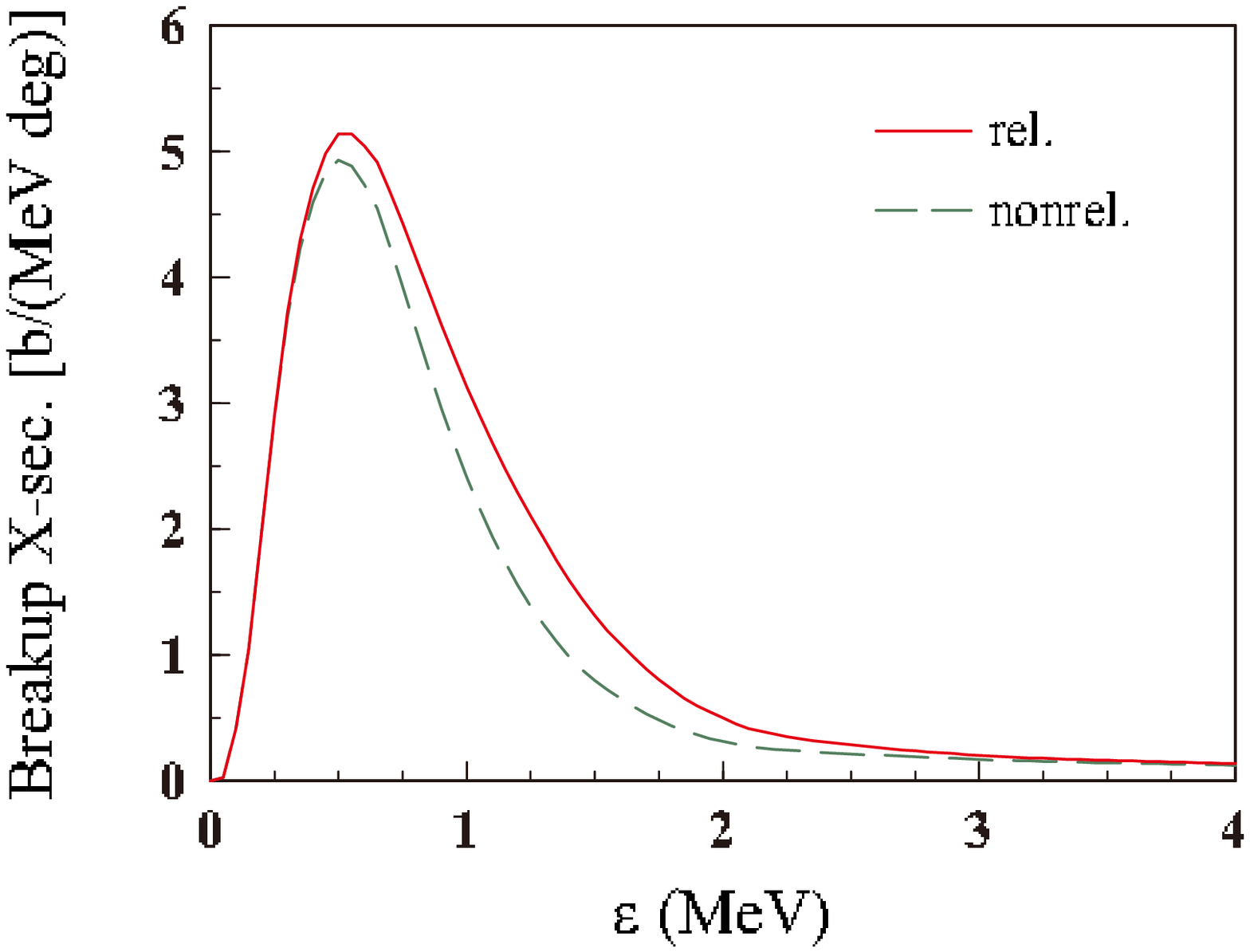}
} \caption{\label{fig2} Comparison between DDBUX, for the
$^8$B$+^{208}$Pb reaction at 250 MeV/nucleon, obtained with
relativistic (solid line) and nonrelativistic (dashed line)
calculations. The left panel shows the DDBUX with $\epsilon=1.5$ MeV
as a function of $\theta$, while the right panel displays the DDX
with $\theta=0.06^\circ$ as a function of $\epsilon$. }
\end{figure}
%
%
\begin{figure}[b]
\centerline{
\includegraphics[width=70mm,keepaspectratio]{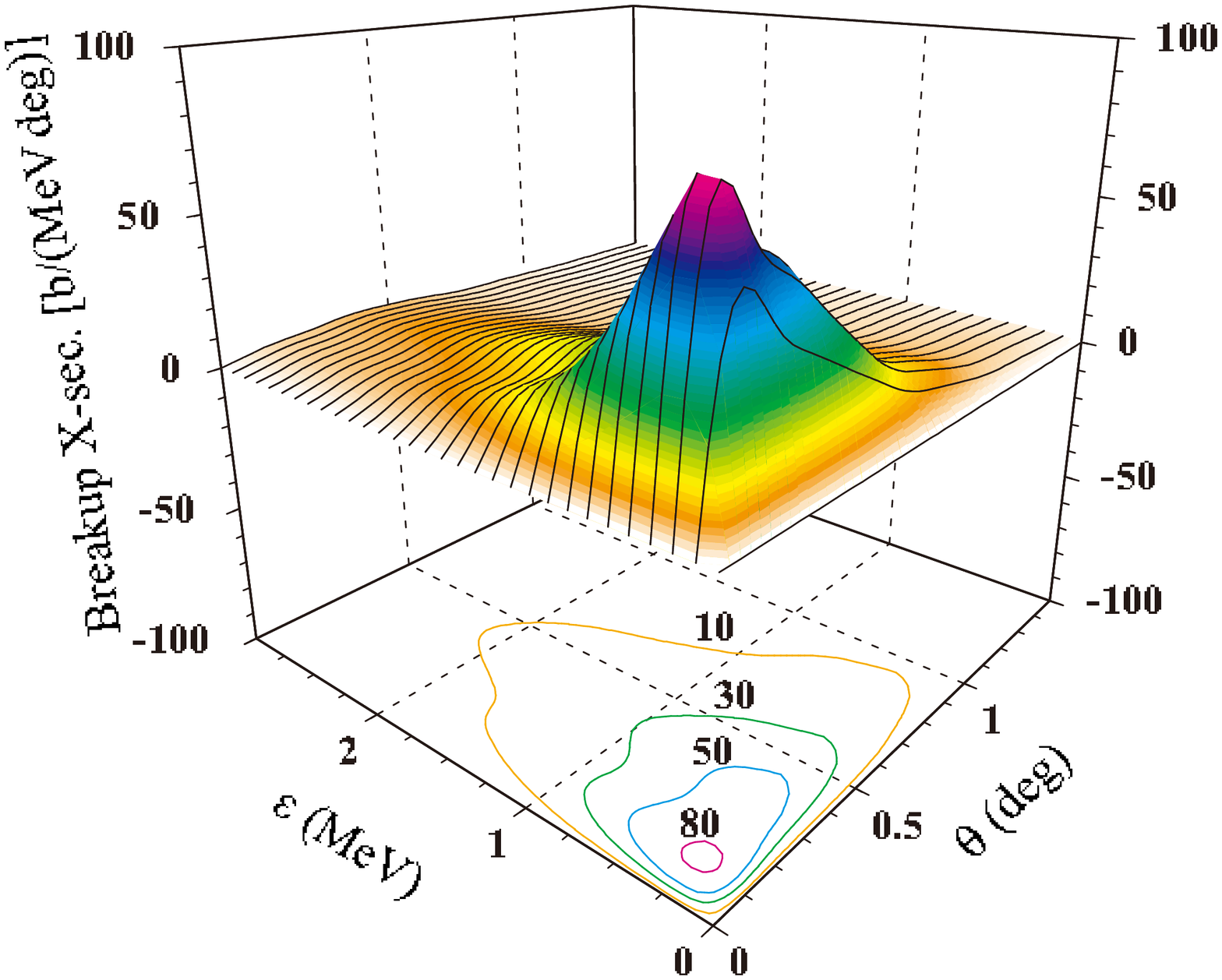}
\includegraphics[width=70mm,keepaspectratio]{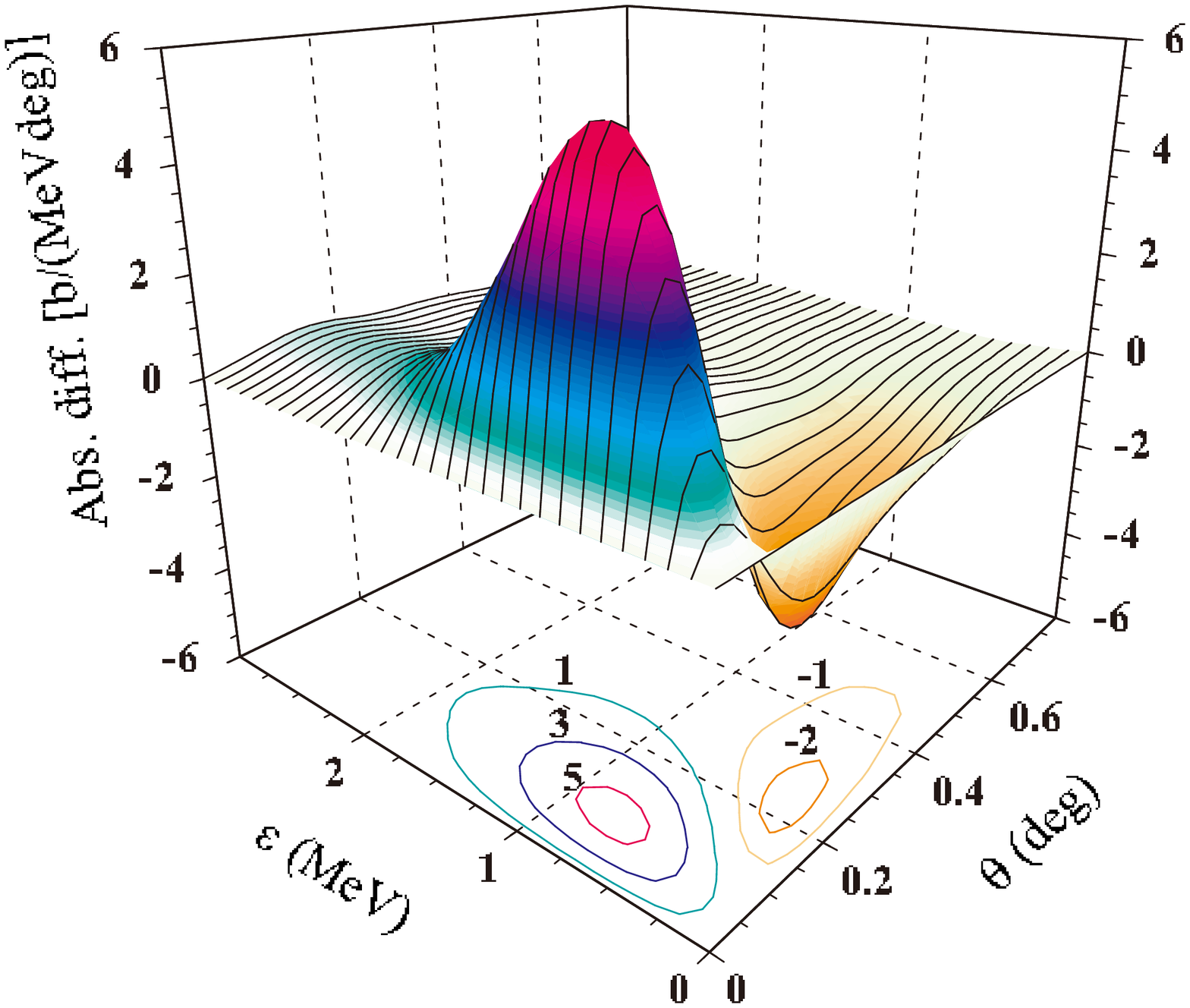}
}
\caption{\label{fig3}
Same as in Fig.~\ref{fig1} but for $^{11}$Be$+^{208}$Pb
at 250 MeV/nucleon.
}
\end{figure}
In the left panel of Fig.~1 we show the double differential breakup
cross section (DDBUX), $d^2 \sigma_{\rm BU}/(d\epsilon d \theta)$,
for the $^8$B$+^{208}$Pb reaction at 250 MeV/nucleon calculated with
relativistic CDCC based on Eq.~\eqref{fh}.
$\epsilon$ is the relative energy of the two
fragments of the projectile after the breakup, and $\theta$ is the
scattering angle of the c.m. of the projectile. The right
panel displays the difference of the DDBUX shown
in the left panel from that
calculated with nonrelativistic CDCC, which also is
based on Eq.~\eqref{fh} but including
no dynamical relativistic corrections.
One observes a rather large
increase in the DDBUX due to relativity at forward angles ($\theta
\la 0.2^\circ$) and around $\epsilon=1$ MeV, where the DDBUX has
quite large values as shown in the left panel. To show this
effect more clearly, we plot the DDBUX with $\epsilon$ ($\theta$)
fixed at 1.5 MeV ($0.06^\circ$) in the left (right) panel of Fig.~2.
The results of the relativistic and nonrelativistic CDCC calculation
are shown by the solid and dashed lines, respectively.

%
\begin{figure}[t]
\centerline{
\includegraphics[width=70mm,keepaspectratio]{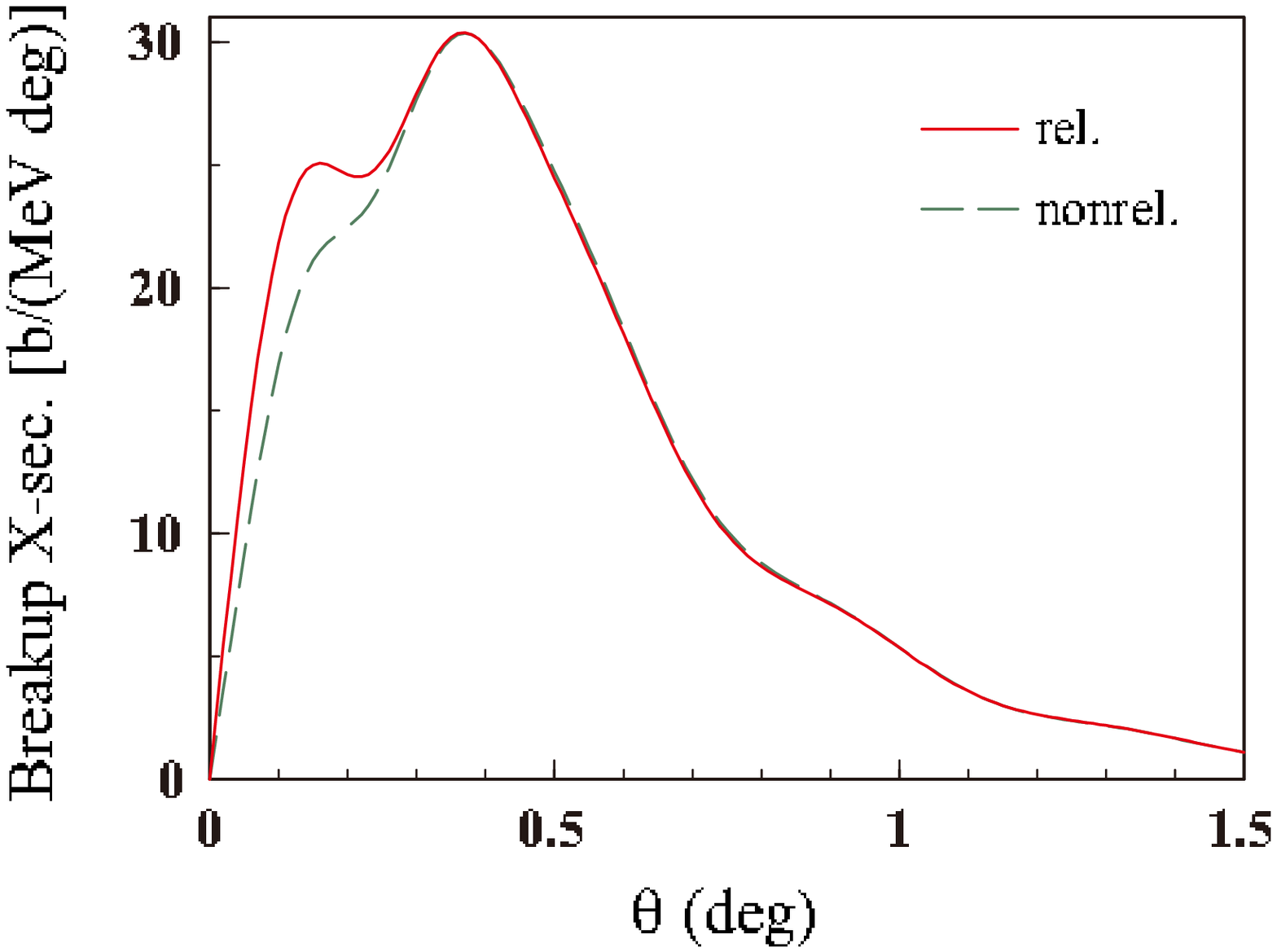}
\includegraphics[width=70mm,keepaspectratio]{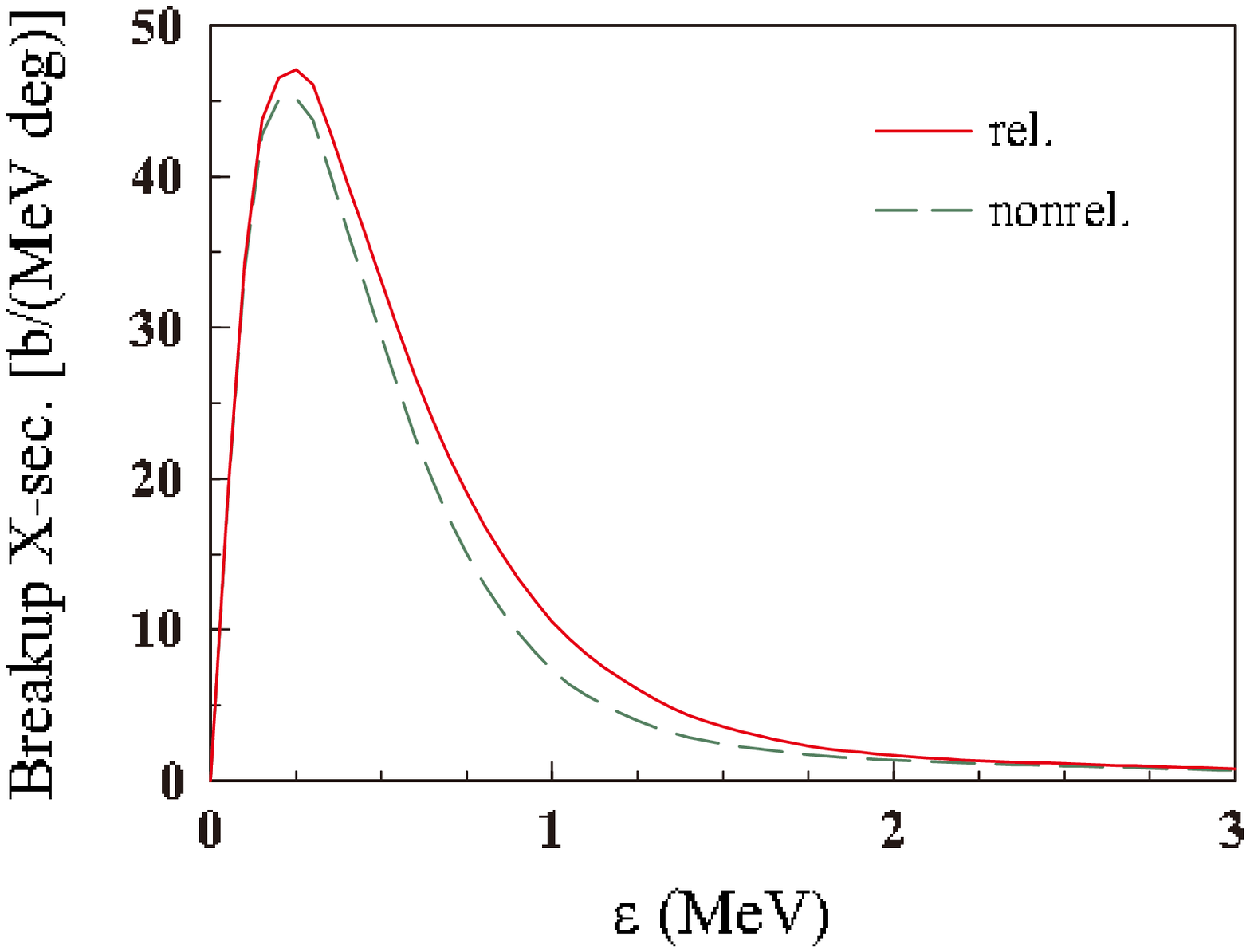}
}
\caption{\label{fig4}
Same as in Fig.~\ref{fig2} but for $^{11}$Be$+^{208}$Pb
at 250 MeV/nucleon;
the left (right) panel corresponds to the DDBUX with
$\epsilon=1.0$ MeV ($\theta=0.04^\circ$) as a function of
$\theta$ ($\epsilon$).
}
\end{figure}
%
%
\begin{figure}[b]
\centerline{
\includegraphics[width=70mm,keepaspectratio]{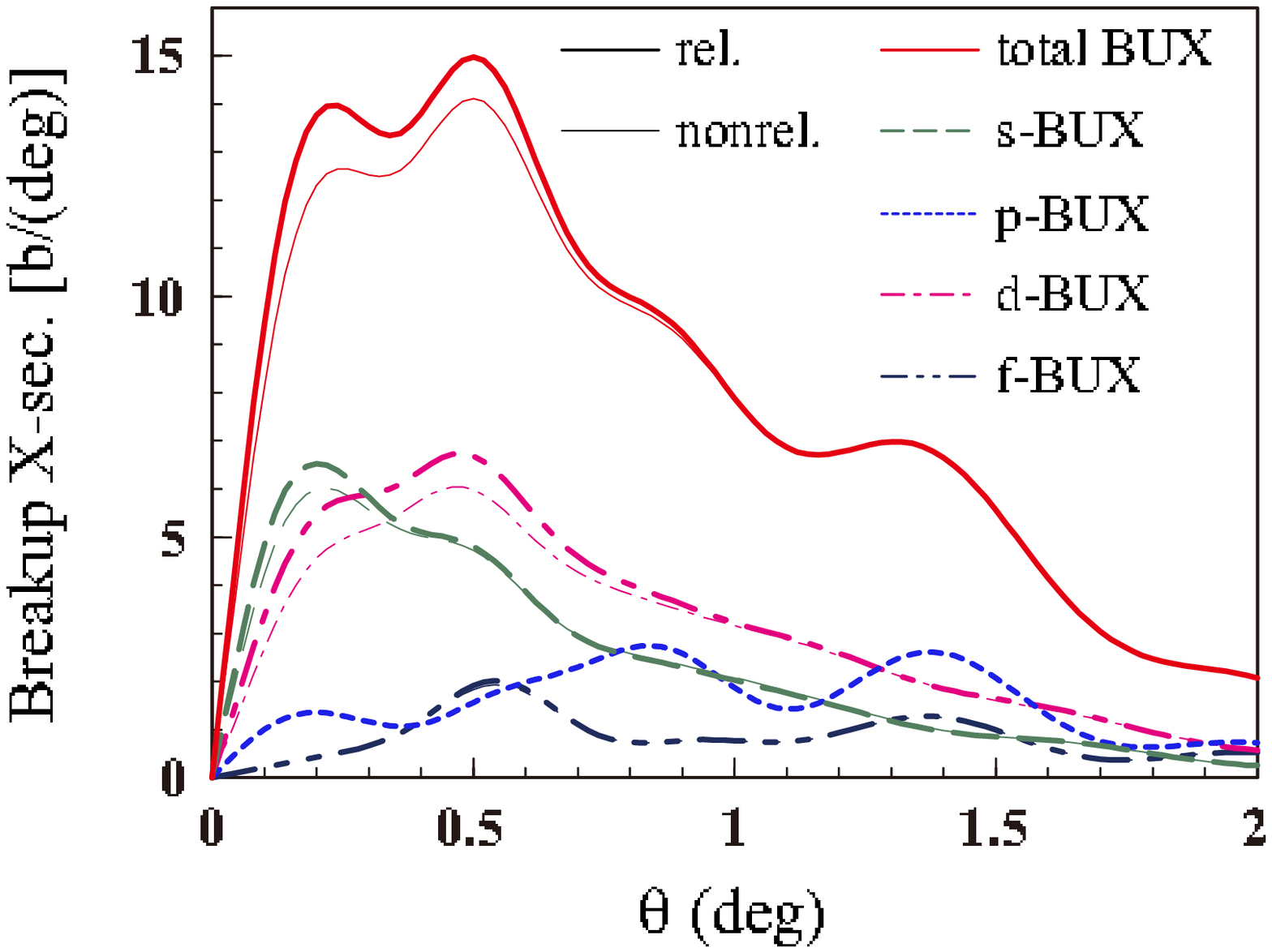}
\includegraphics[width=70mm,keepaspectratio]{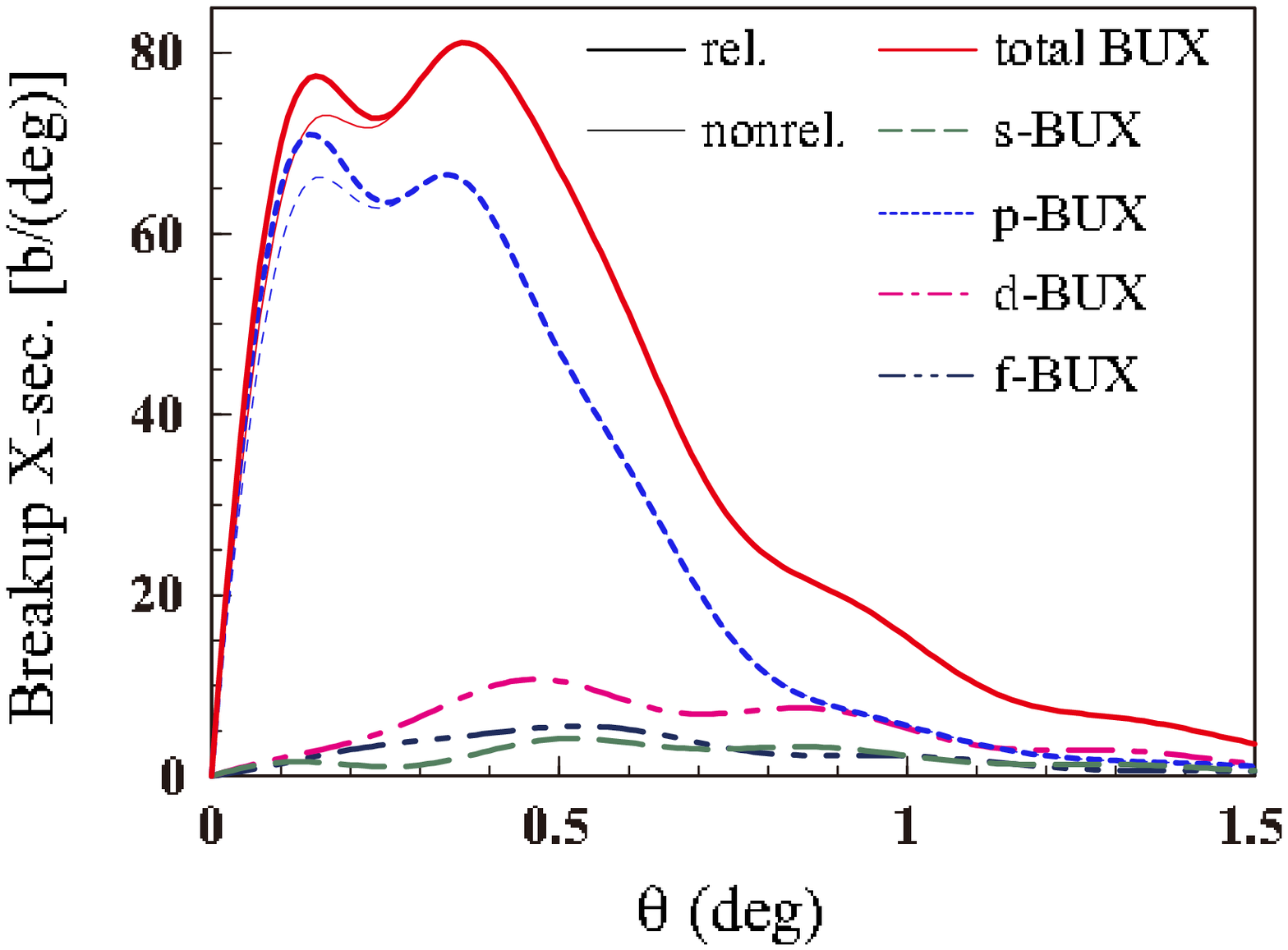}
} \caption{\label{fig5} Angular distribution of the breakup cross
sections (BUX) to the s-, p-, d-, and f-states of the projectile are
shown by the dashed, dotted, dash-dotted, and dash-dot-dotted lines,
respectively, and the solid line represents the sum of them. The
thick (thin) lines are the results of the relativistic
(nonrelativistic) calculation. The left and right panels correspond
to the $^8$B$+^{208}$Pb and $^{11}$Be$+^{208}$Pb reactions,
respectively. }
\end{figure}
The results for the $^{11}$Be$+^{208}$Pb reaction at 250 MeV/nucleon
are shown in Figs.~3 and 4, in the same way as in Figs.~1 and 2,
respectively. Features of the results are very similar to those of
the $^{8}$B$+^{208}$Pb reaction, except that 1) the magnitude of the
DDBUX is much larger than that of $^{8}$B and 2) the increase in the
DDBUX due to relativity is slightly smaller and limited at smaller
angles. The larger magnitude of the DDBUX arises from the larger
value of the E1 effective charge $e_{{\rm E}1}$ of $^{11}$Be
compared with that of $^8$B. The smaller effect of relativity for
this reaction can be understood by the decomposition of the breakup
cross section (BUX) into the components corresponding to individual
partial waves of the projectile.

In the left (right) panel of Fig.~5 we show the BUX to the s-, p-,
d-, and f-waves of $^8$B ($^{11}$Be) by the dashed, dotted,
dash-dotted, and dash-dot-dotted lines, respectively; the solid line
is the sum of them. Note that the cross sections are obtained
by integrating the DDBUX over $\epsilon$.
In each panel, the thick and thin lines
represent the results of relativistic and nonrelativistic
CDCC, respectively.
Since the ground state of $^8$B is the p-wave, s- and
d-state BUX are dominant in the breakup of $^8$B, as the E1
transition gives the dominant contribution to the breakup reactions
caused by a heavy target nucleus. For the breakup of $^{11}$Be, the
ground state of which is assumed here to be an s-wave, the dominant
contribution is to the p-wave. One finds the increase of BUX due to
relativity only in the p-wave channel, corresponding to the E1
transitions. The increase (due to relativity) in the s-wave BUX of
$^8$B has the same feature as that of the p-wave BUX of $^{11}$Be.
For the $^8$B breakup, the increase in the d-wave BUX is also
noticeable, which makes the relativistic effects on the total BUX of
$^8$B more significant.

%
\begin{figure}[t]
\centerline{
\includegraphics[width=70mm,keepaspectratio]{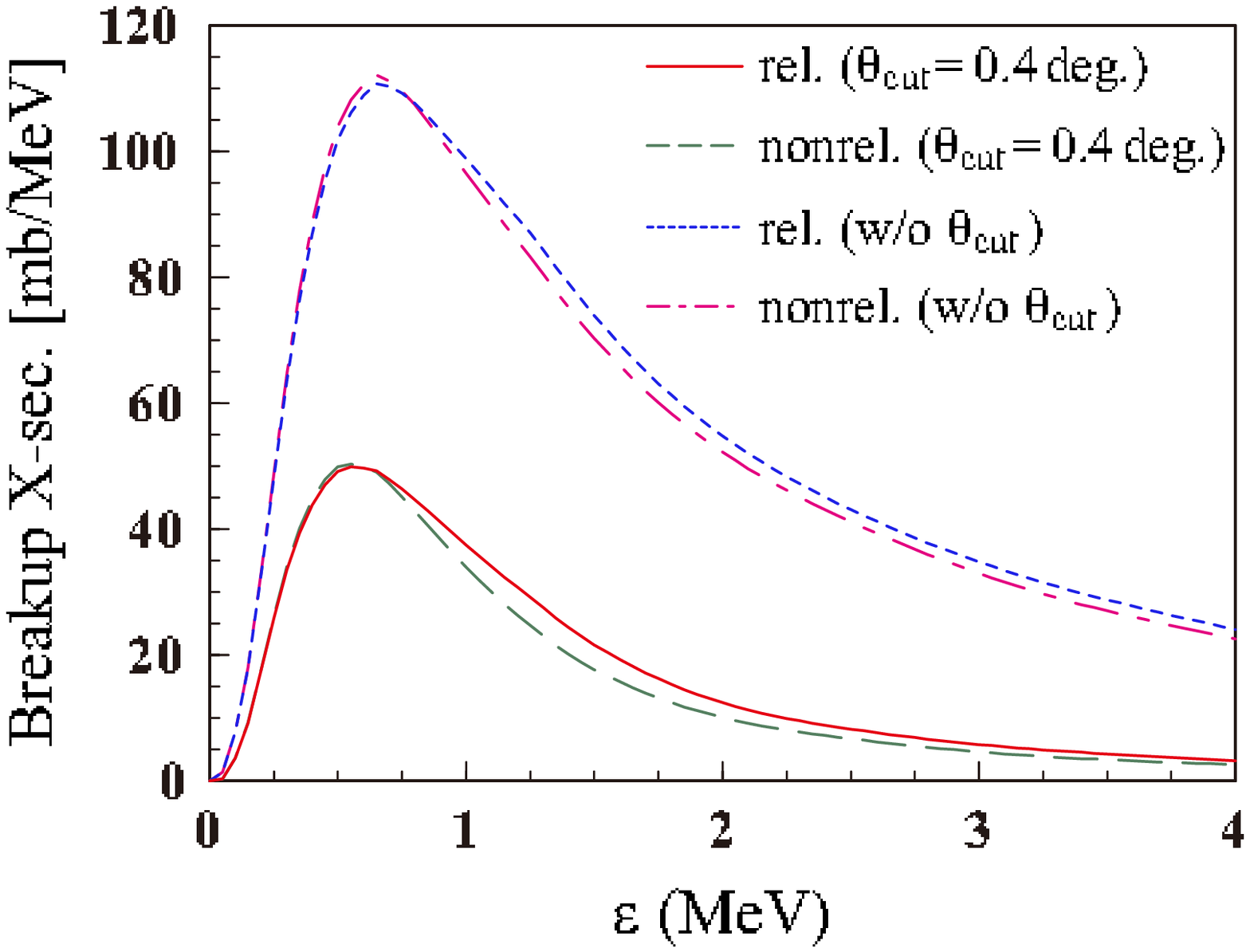}
\includegraphics[width=70mm,keepaspectratio]{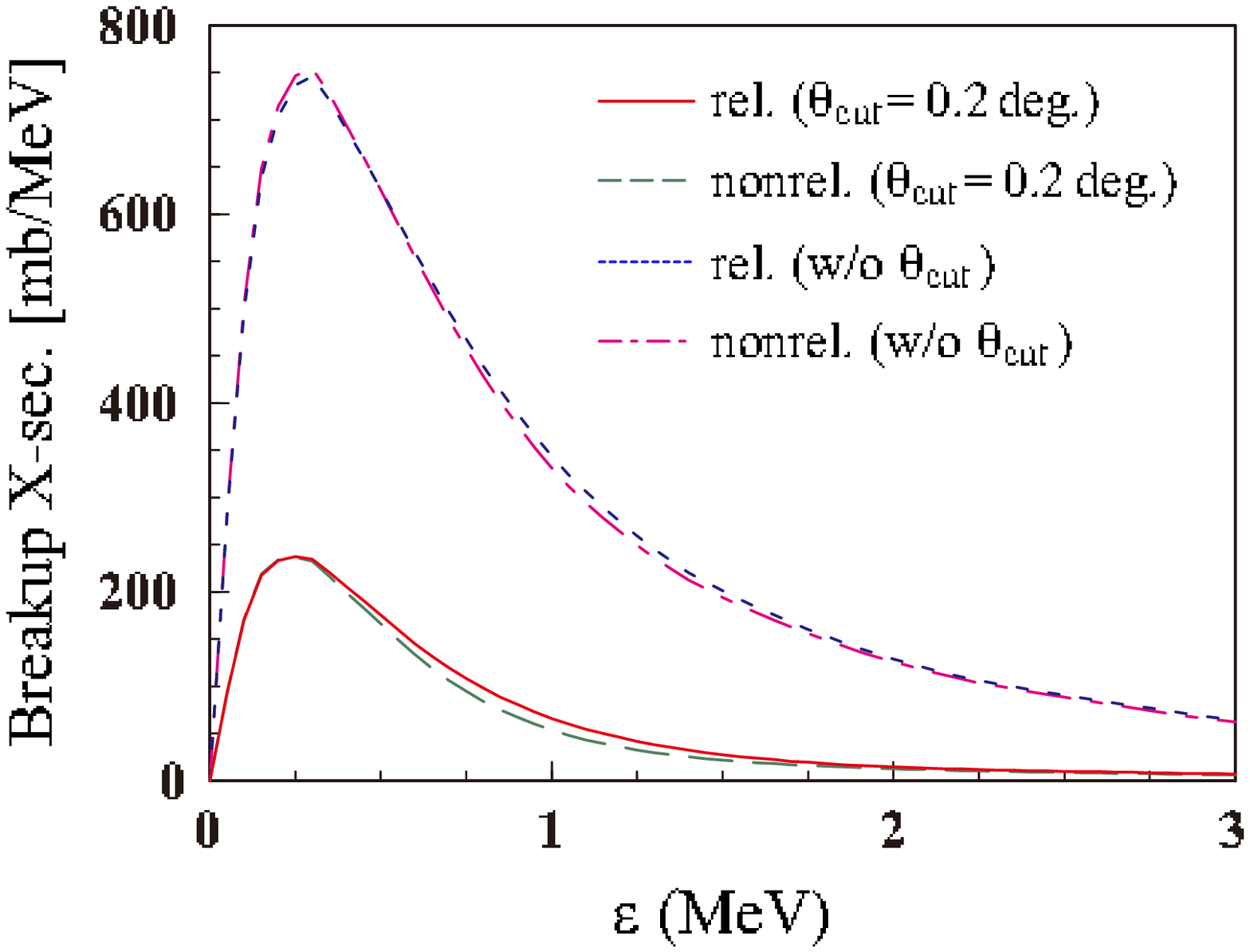}
} \caption{\label{fig6} Breakup energy spectra for $^8$B$+^{208}$Pb
at 250 MeV/nucleon (left panel). The solid and dashed lines show the
results of the relativistic and nonrelativistic calculations,
respectively, with a maximum c.m. scattering angle, $\theta_{\rm
max}$, equal to 0.4$^\circ$. The dotted (dash-dotted) line is the
same as the solid (dashed) line but $\theta_{\rm max}=5.0^\circ$.
Results for $^{11}$Be$+^{208}$Pb at 250 MeV/nucleon are shown in the
right panel. Here, $\theta_{\rm max}=0.2^\circ$ ($3.0^\circ$) for
the solid and dashed (dotted and dash-dotted) lines. }
\end{figure}
In the analysis of a breakup experiment using a heavy target, a
cutoff value for $\theta$, $\theta_{\rm cut}$, is usually introduced
which aims to eliminate contributions from nuclear-induced breakup.
Since the increase in the DDBUX due to relativity is located at
forward angles, such an analysis with $\theta_{\rm cut}$ requires a
relativistic description of the breakup reaction. This is clearly
shown in Fig.~6, where the $\theta_{\rm cut}$ dependence of the
relativistic effects on the breakup energy spectrum is plotted. The
left and right panels correspond to the $^{8}$B and $^{11}$Be
breakup reactions, respectively. The solid (dotted) and dashed
(dash-dotted) lines represent the results of
relativistic and nonrelativistic
CDCC with (without) $\theta_{\rm cut}$, respectively. Here,
$\theta_{\rm cut}=0.4^\circ$ ($0.2^\circ$) is used for the $^{8}$B
($^{11}$Be) breakup. One sees that the use of a cutoff angle
$\theta_{\rm cut}$ enhances the relativistic effects. Note that
preceding analyses of Coulomb breakup processes using the virtual
photon method, or first-order perturbation theory, included
relativistic effects properly. They would give, however, very
different results from those of relativistic CDCC, because of the
contributions of nuclear breakup and higher-order processes in
Coulomb breakup, as shown in \S\ref{sec3-3}.

%
\begin{figure}[t]
\centerline{
\includegraphics[width=70mm,keepaspectratio]{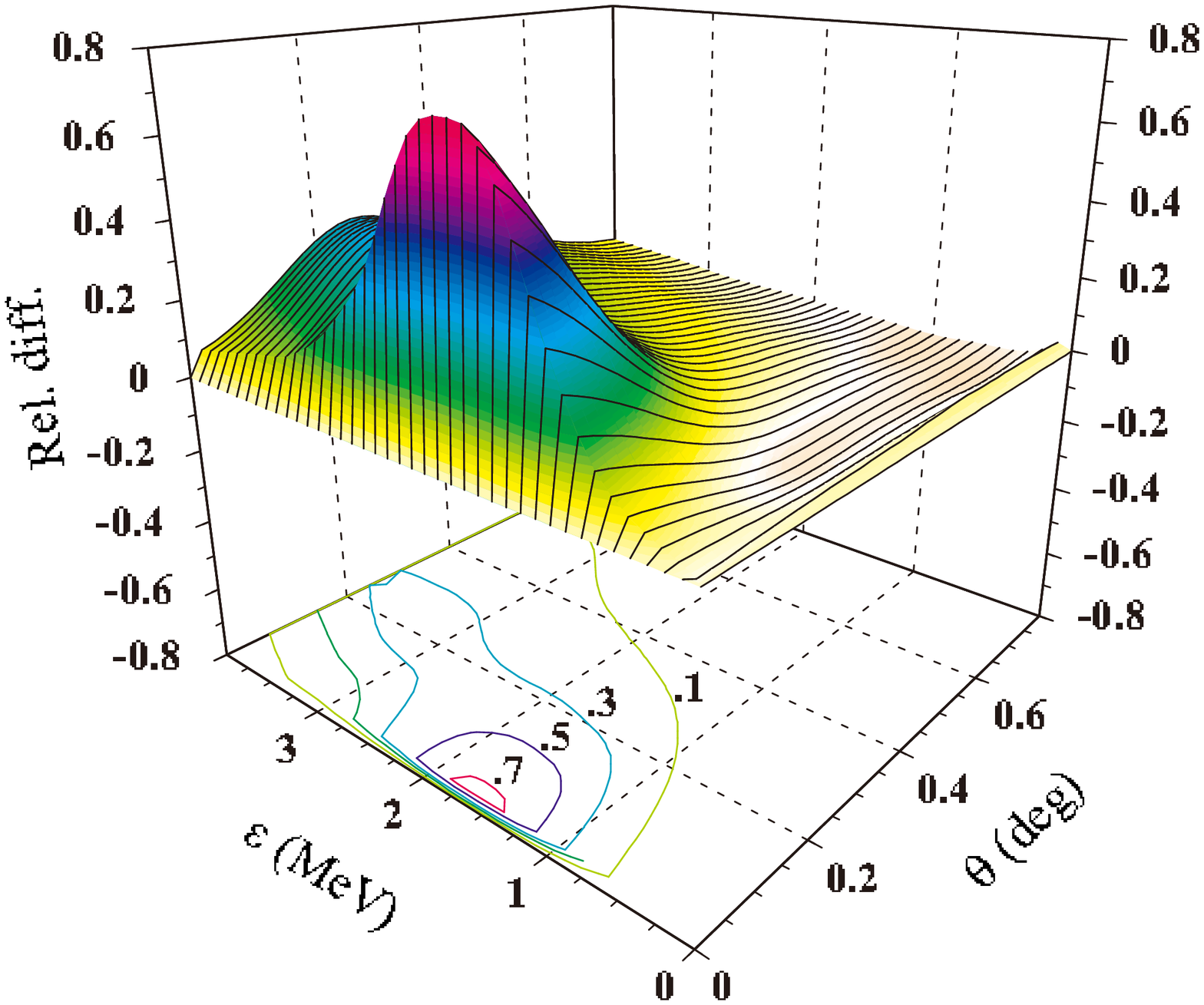}
\includegraphics[width=70mm,keepaspectratio]{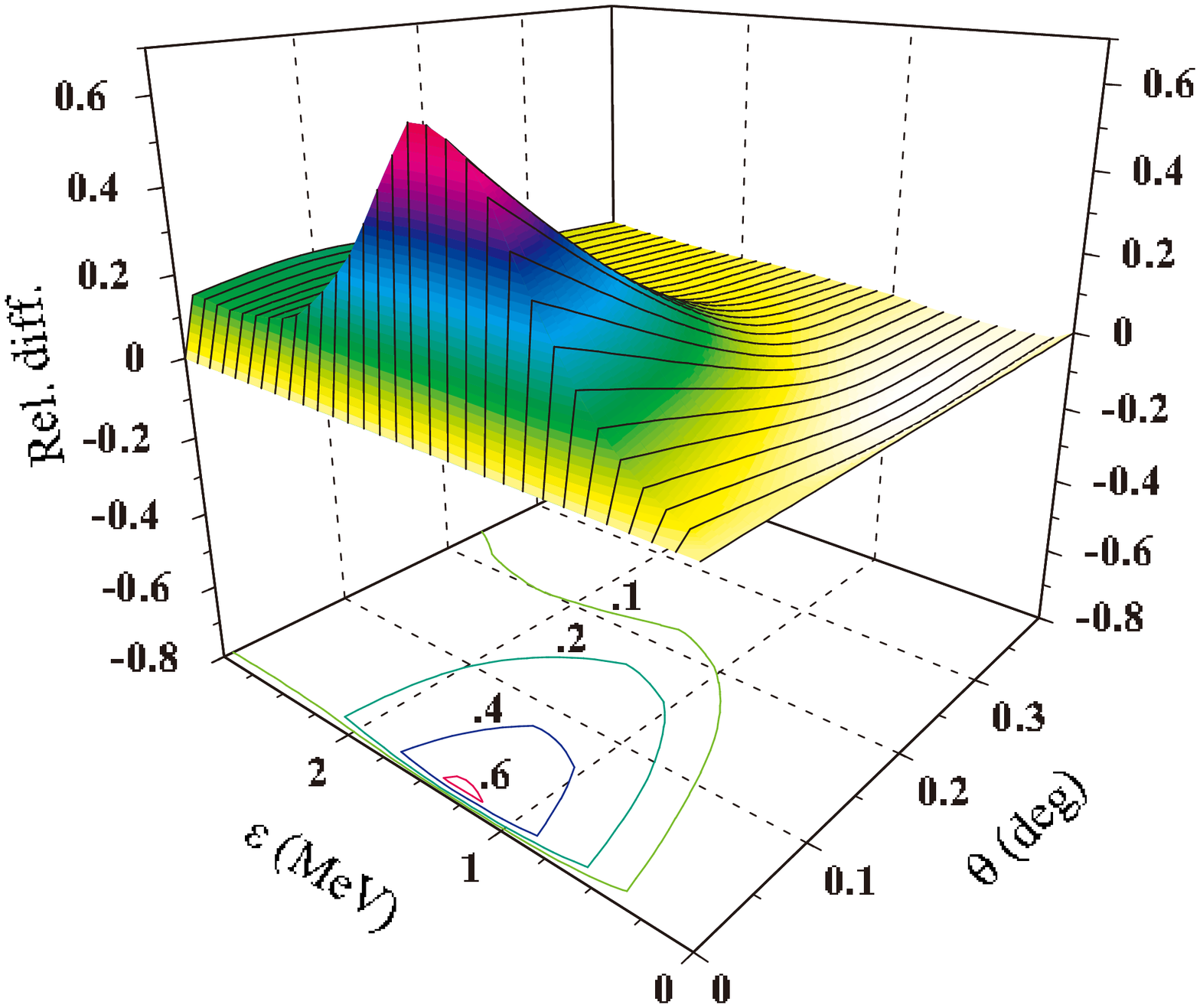}
}
\caption{\label{fig7}
Relative difference of the relativistic DDX from the
nonrelativistic one.
The left and right panels respectively correspond to
$^8$B$+^{208}$Pb and $^{11}$Be$+^{208}$Pb at 250 MeV/nucleon.
}
\end{figure}
%
%
\begin{figure}[b]
\centerline{
\includegraphics[width=70mm,keepaspectratio]{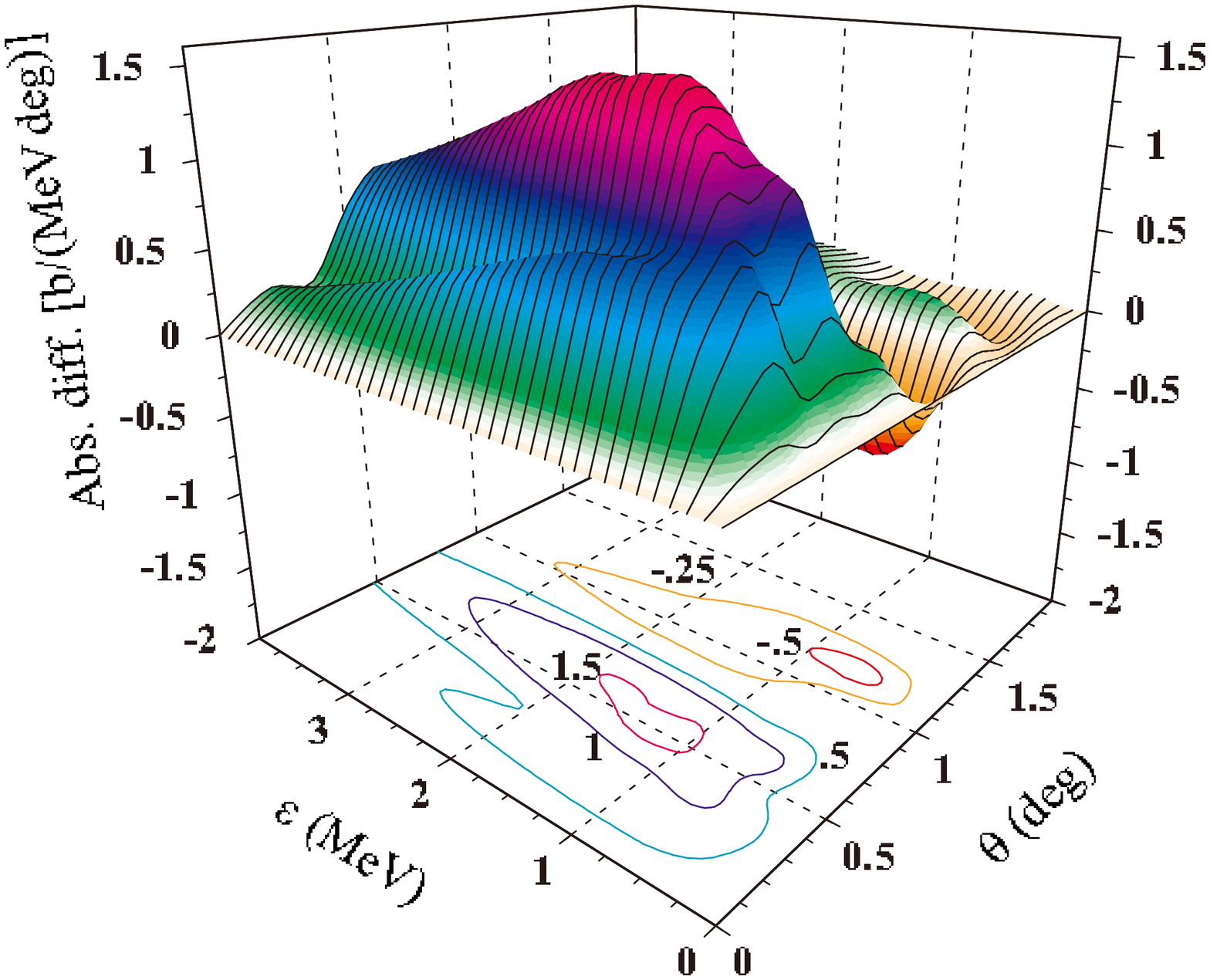}
\includegraphics[width=70mm,keepaspectratio]{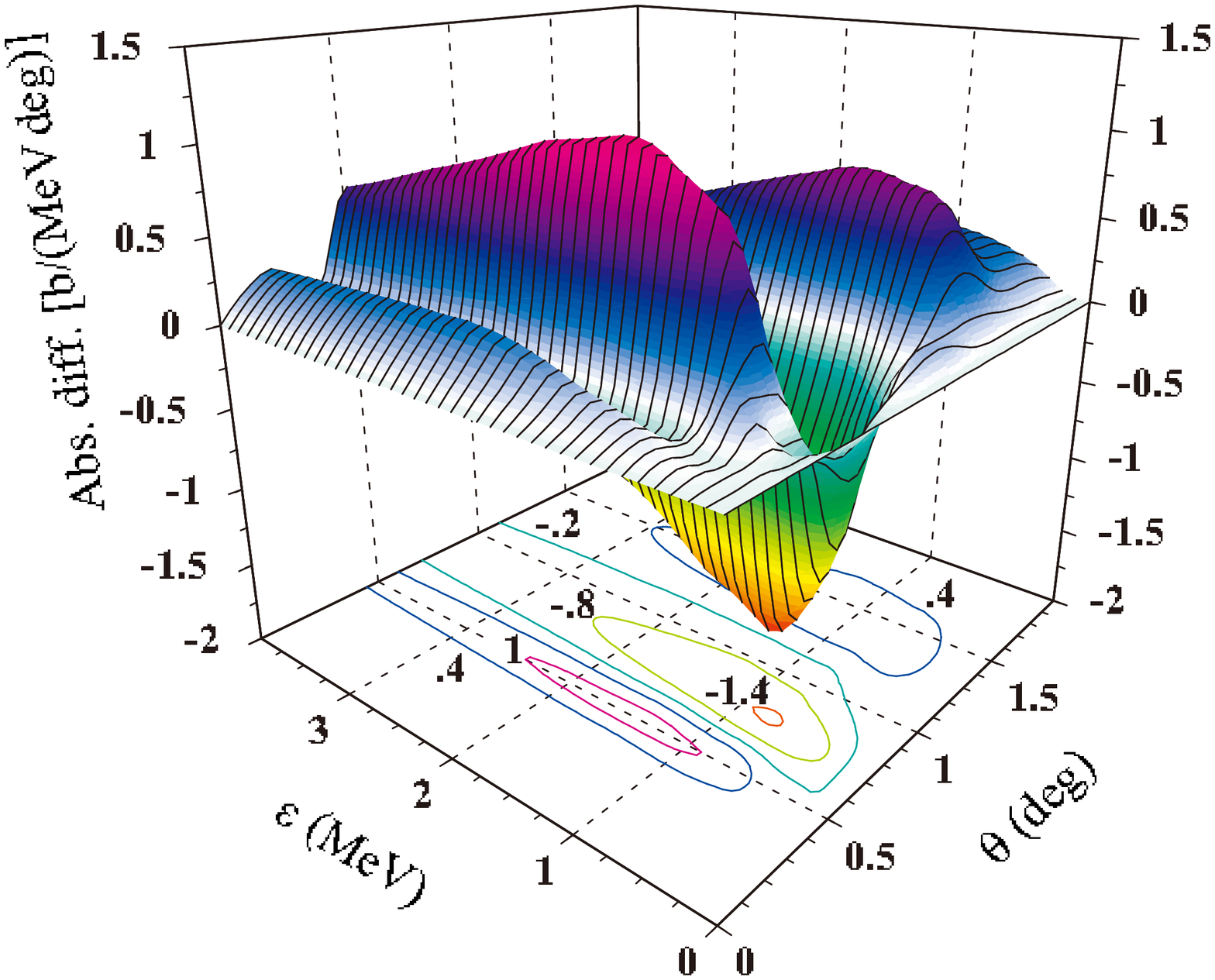}
} \caption{\label{fig8} The left panel shows the difference of the
DDBUX for $^8$B$+^{208}$Pb at 250 MeV/nucleon obtained with
a first-order calculation from
that obtained with CDCC. The right panel shows
the difference of the DDX obtained with CDCC including nuclear and
Coulomb breakup to that with only Coulomb breakup. In all
calculations we take into account dynamical relativistic
corrections. }
\end{figure}
Our conclusions based on the previous figures are better seen if we
plot relative differences of the several effects we want to discuss
here. For example, we show in the left (right) panel of Fig.~7 the
relative difference of the DDBUX obtained with relativistic CDCC
from that with nonrelativistic CDCC, for the $^{8}$B ($^{11}$Be) breakup. The
difference is indeed large (several tens of \% level),
which shows the importance of
the relativistic effects on the DDBUX.

\subsection{Higher-order effects and nuclear breakup}
\label{sec3-3}

%
\begin{figure}[b]
\centerline{
\includegraphics[width=70mm,keepaspectratio]{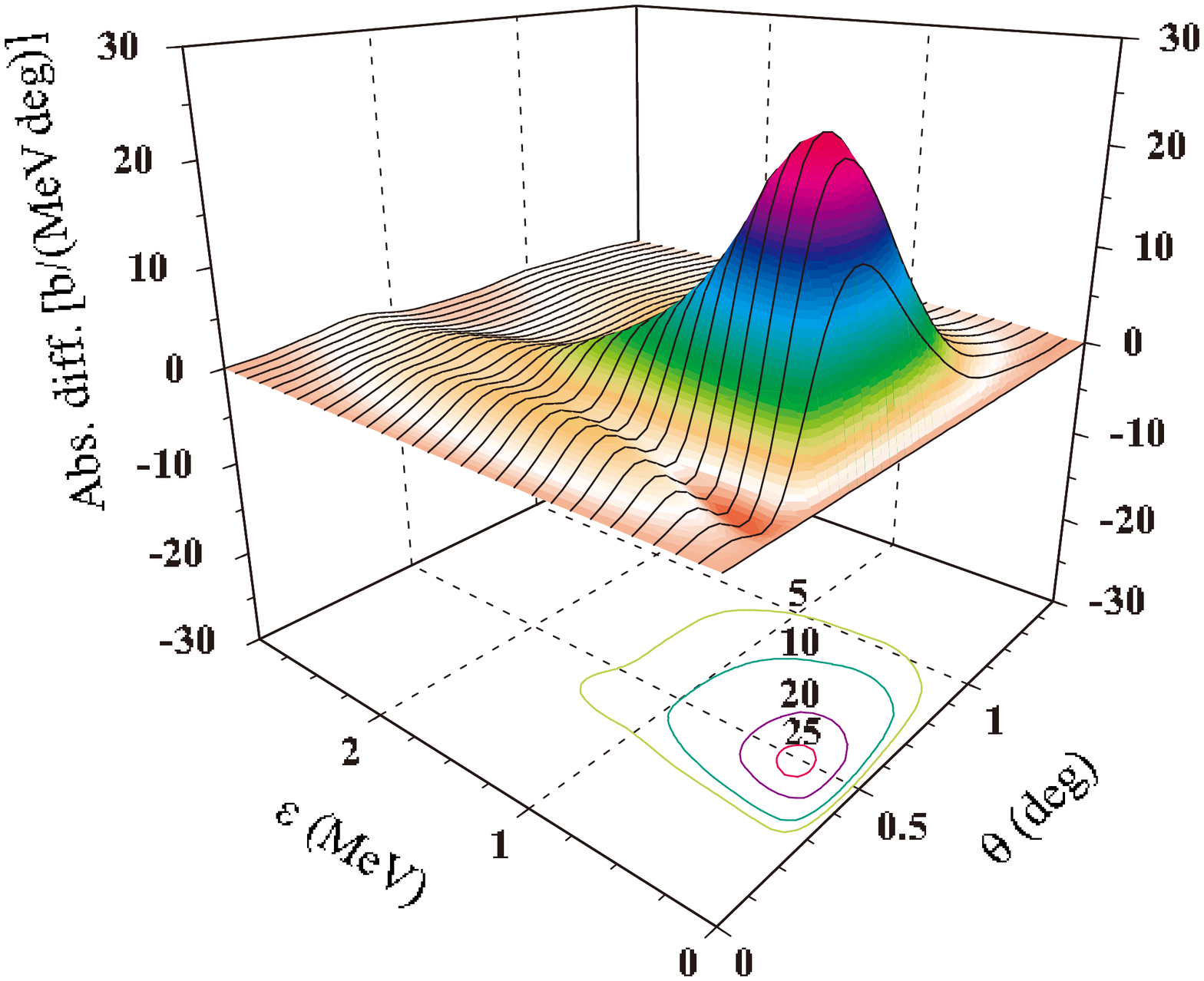}
\includegraphics[width=70mm,keepaspectratio]{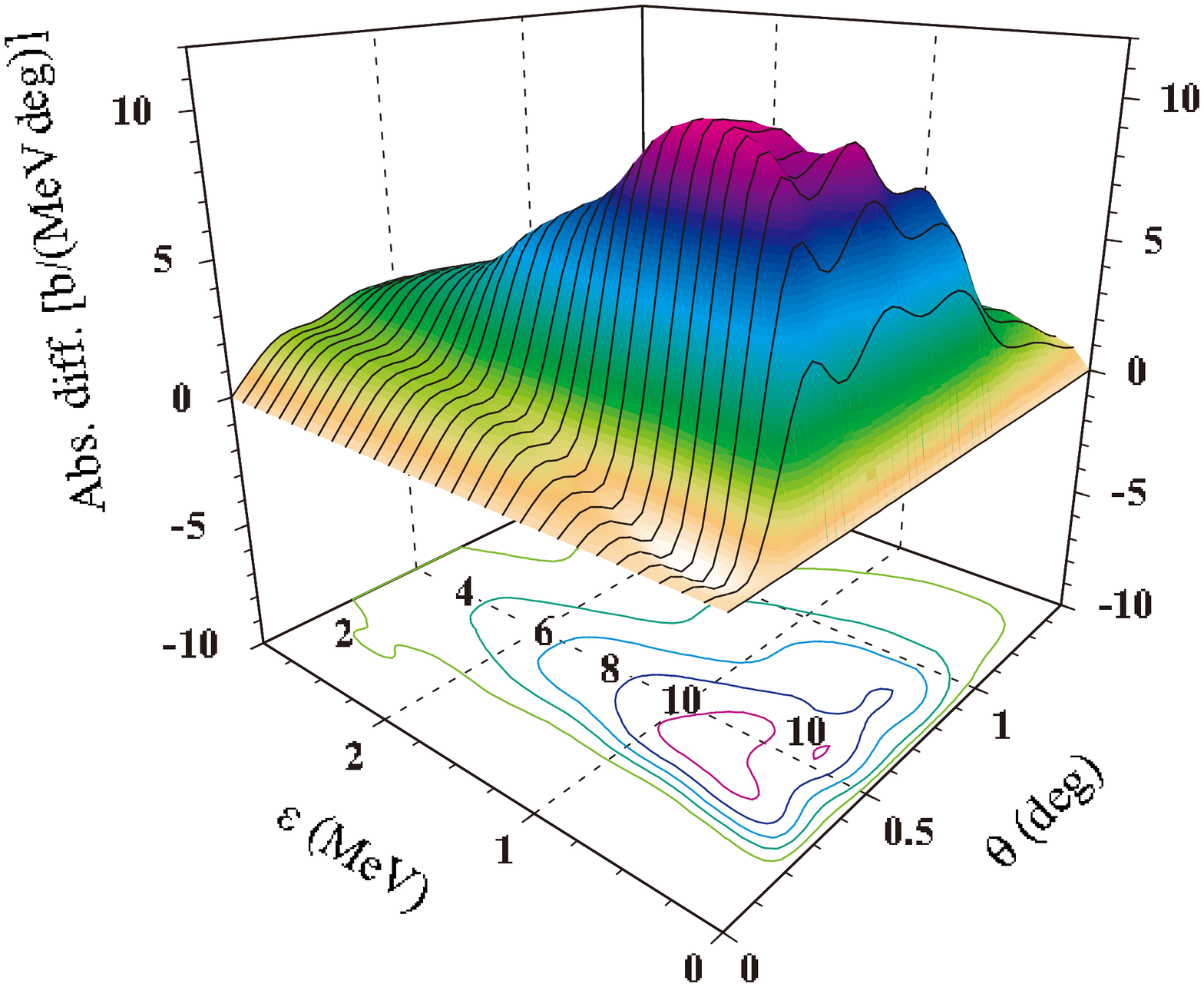}
}
\caption{\label{fig9}
Same as in Fig.~\ref{fig7} but for $^{11}$Be$+^{208}$Pb
at 250 MeV/nucleon.
}
\end{figure}

As shown in Ref.~\citen{OB09}, the contribution of nuclear breakup
as well as higher-order effects on the BUX is very important, even
at 250 MeV/nucleon. We show in the left panel of Fig.~8 the absolute
difference of the DDBUX for $^8$B$+^{208}$Pb calculated with first
order perturbation theory from that with relativistic CDCC,
which we call
here {\it higher-order corrections}. The right panel shows the
difference of the DDBUX obtained with relativistic CDCC,
which includes both nuclear and Coulomb breakup,
from that including only the Coulomb breakup. We
call this difference {\it nuclear breakup correction}; the effects of the
interference between the nuclear and Coulomb breakup amplitudes
are also included. One
sees the higher-order and nuclear breakup corrections are of the
order of 10\%. They have a rather weak $\epsilon$ dependence, while
they oscillate with respect to $\theta$.

As shown in the left panel of Fig.~5, all partial-wave components of
the BUX of $^8$B are comparable at $\theta \ga 0.5^\circ$. This feature
results in a non-trivial change in each BUX component due to the
inclusion of higher-order processes or nuclear breakup. On the other
hand, these corrections for $^{11}$Be breakup, as shown in Fig.~9,
display a simpler distribution on the $\epsilon$-$\theta$ plane. We
find from a detailed analysis that since the p-state BUX is dominant
for the $^{11}$Be breakup, the change in the p-state BUX due to the
inclusion of the higher-order processes or nuclear breakup and
that in other small BUX components add up almost incoherently.

\subsection{Quantum mechanical effects}
\label{sec3-4}

%
\begin{figure}[t]
\centerline{
\includegraphics[width=70mm,keepaspectratio]{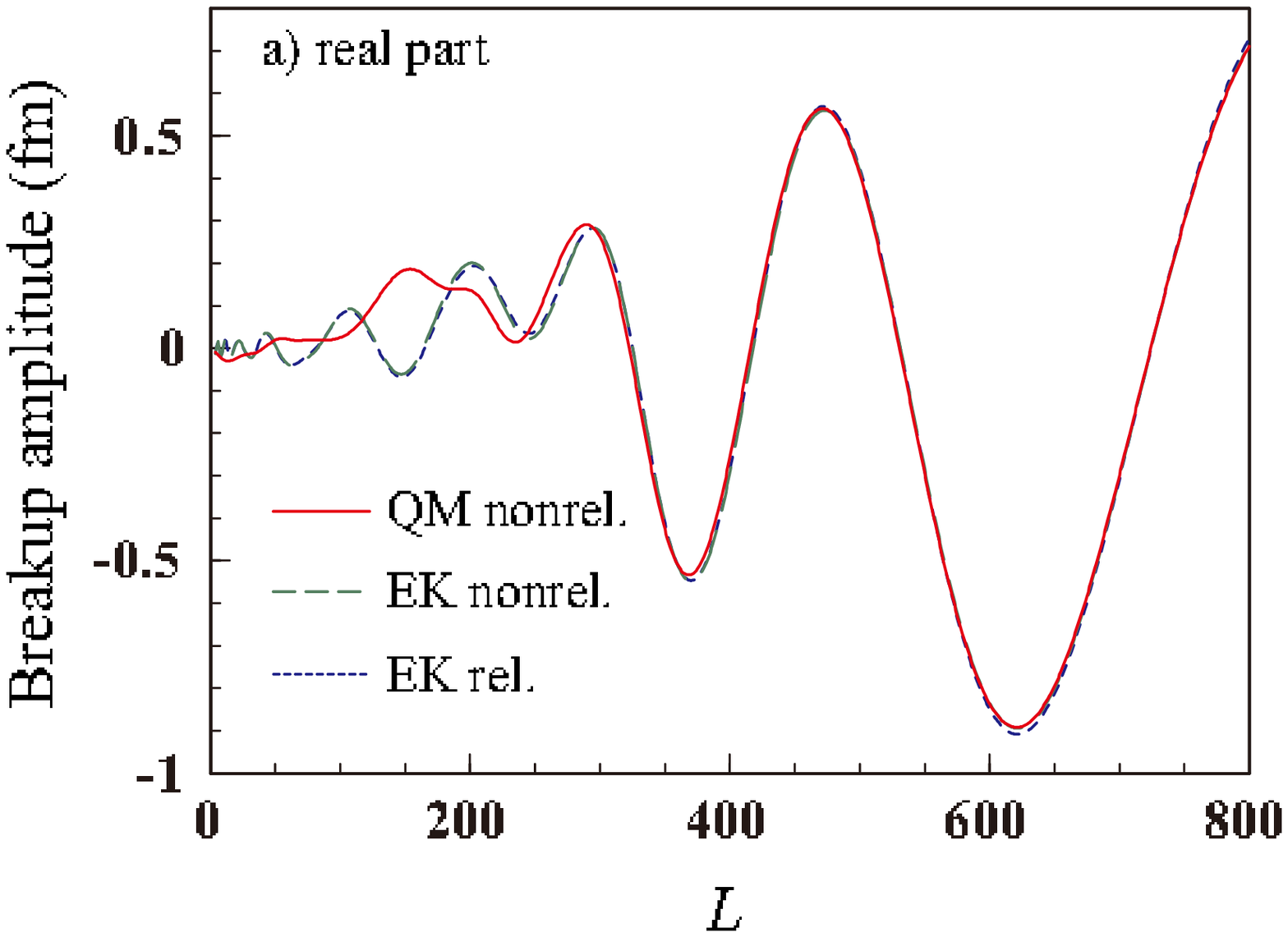}
\includegraphics[width=70mm,keepaspectratio]{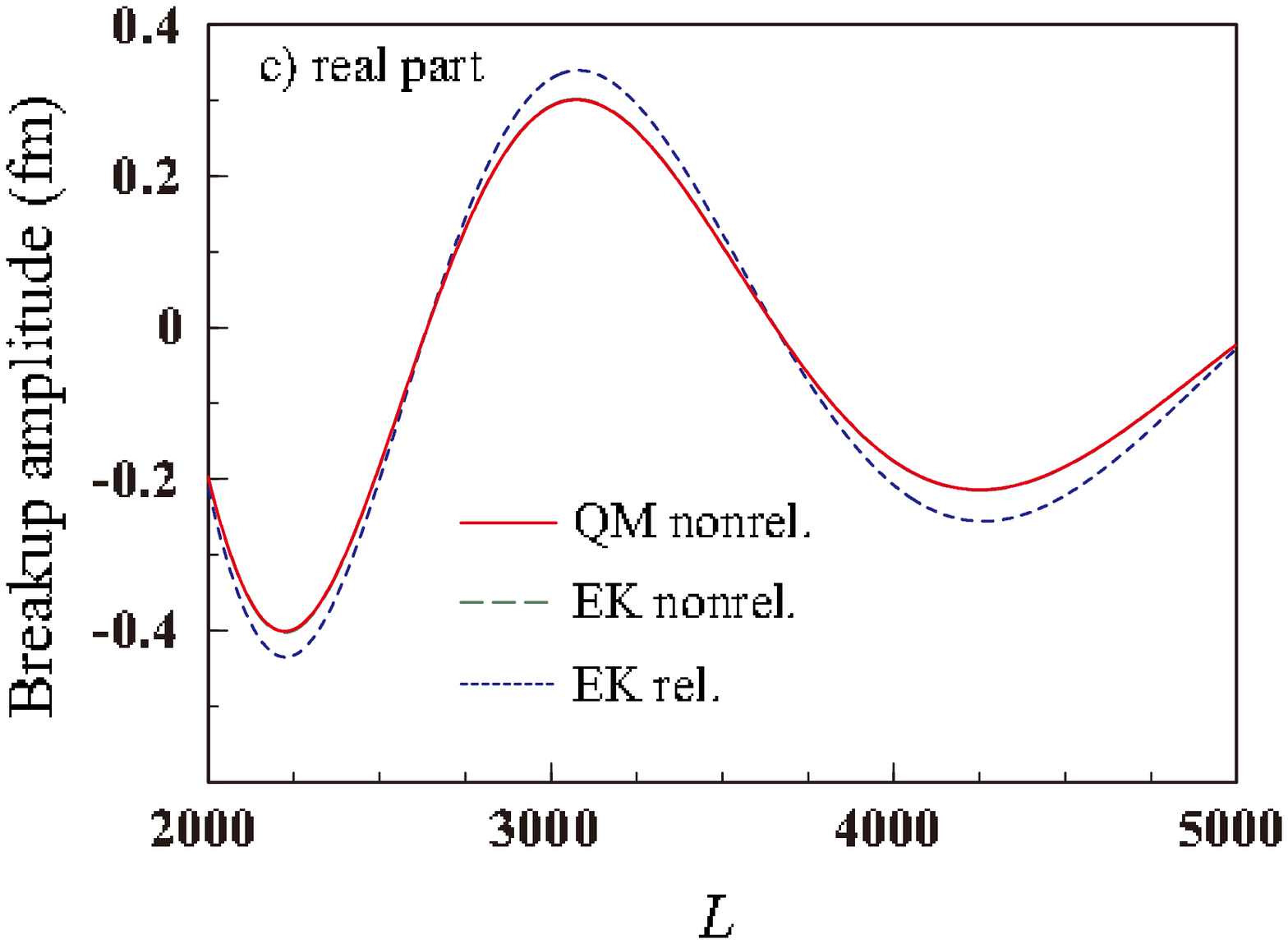}
}
\centerline{
\includegraphics[width=70mm,keepaspectratio]{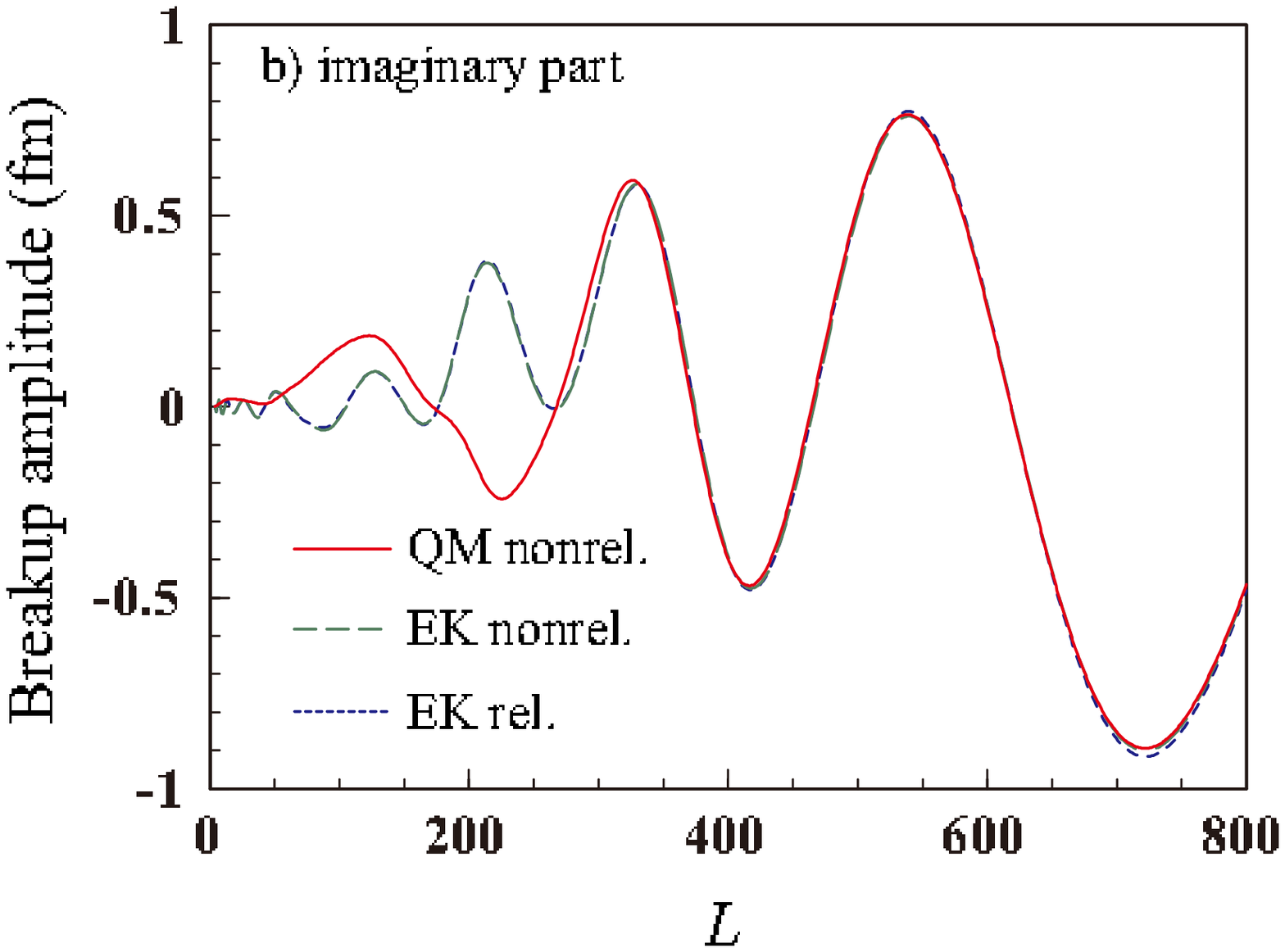}
\includegraphics[width=70mm,keepaspectratio]{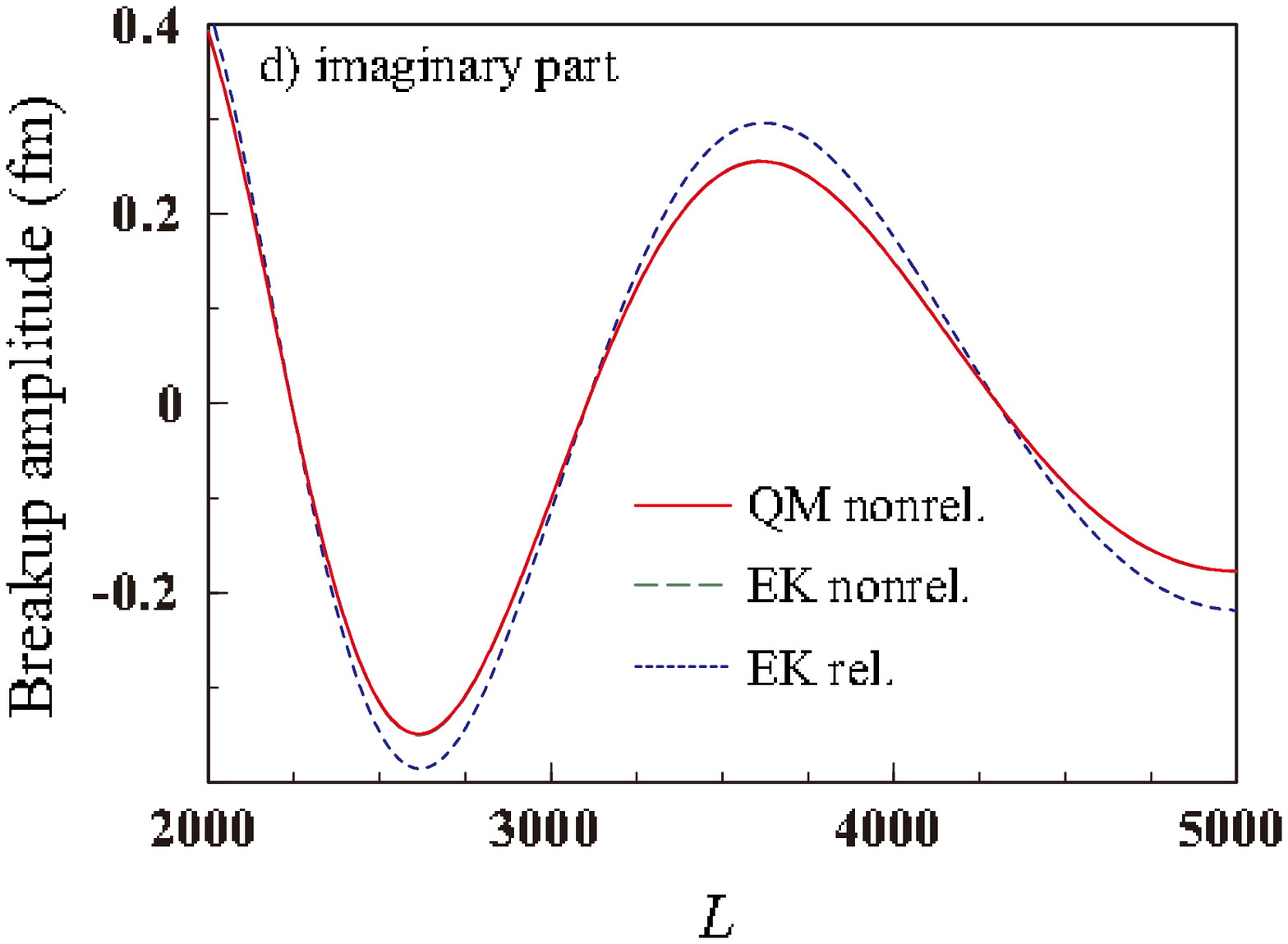}
} \caption{\label{fig10} Breakup amplitude for $^8$B$+^{208}$Pb at
250 MeV/nucleon as a function of the orbital angular momentum $L$
between $^8$B and $^{208}$Pb. The final state of $^8$B is chosen to
be the s-wave 6th bin state, and the $z$-component $m_0$ of the spin
of $^8$B in the incident channel chosen as 1. The solid, dashed, and
dotted lines show the results of nonrelativistic quantum mechanical
CDCC, nonrelativistic eikonal CDCC, and relativistic eikonal CDCC,
respectively. The upper (lower) panels correspond to the real and
imaginary parts of the breakup amplitude. }
\end{figure}
%
%
\begin{figure}[t]
\centerline{
\includegraphics[width=70mm,keepaspectratio]{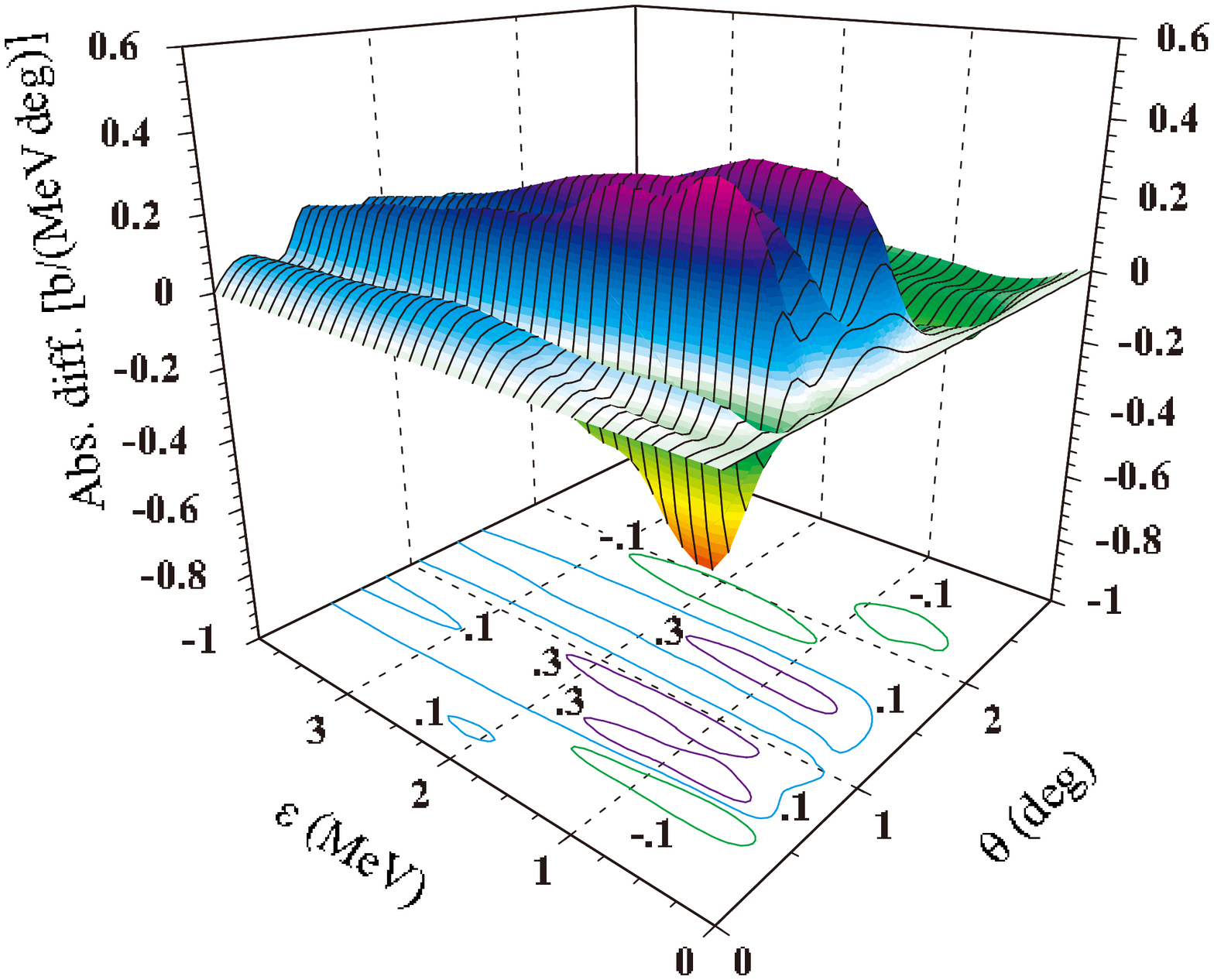}
\includegraphics[width=70mm,keepaspectratio]{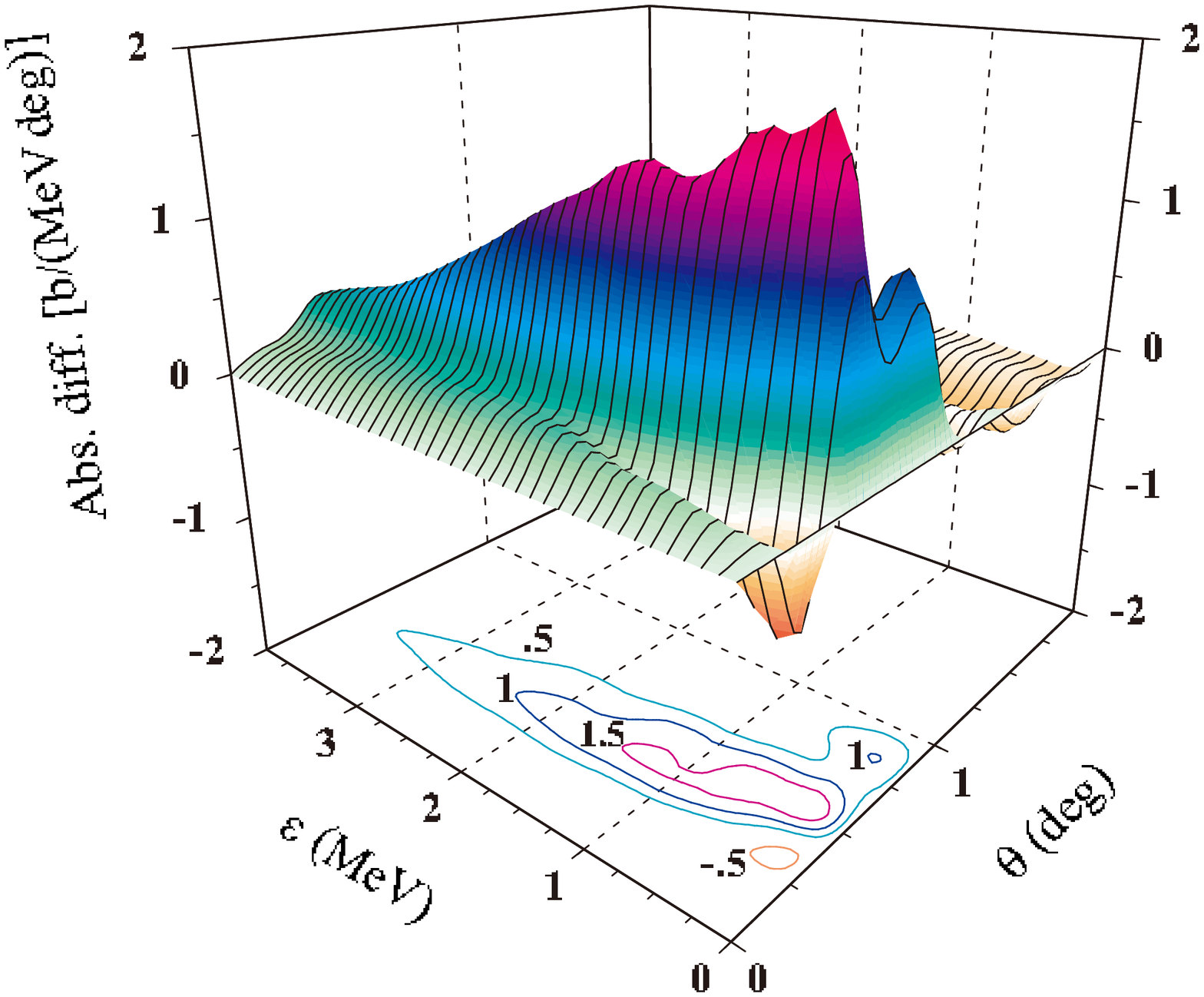}
} \caption{\label{fig11} Difference of the DDX obtained with
 nonrelativistic
quantum-mechanical CDCC from that obtained with
nonrelativistic eikonal CDCC. The left
(right) panel corresponds to $^8$B$+^{208}$Pb ($^{11}$Be$+^{208}$Pb)
at 250 MeV/nucleon. }
\end{figure}
An important feature of the present CDCC calculation is that the
quantum mechanical (QM) correction is explicitly included. This is
possible because, as discussed in Ref.~\citen{OB09}, relativistic
corrections in the continuum-continuum couplings are only
appreciable for the breakup amplitudes corresponding to large values
of $L$, i.e., for large orbital angular momenta of relative motion
between the projectile and target, where the QM correction is
negligibly small~\cite{Og03,Og06}. Note that inclusion of the
relativistic Coulomb and nuclear coupling potentials in a QM
calculation, based on a conventional partial wave decomposition, is
very complicated.
%
\begin{figure}[b]
\centerline{
\includegraphics[width=70mm,keepaspectratio]{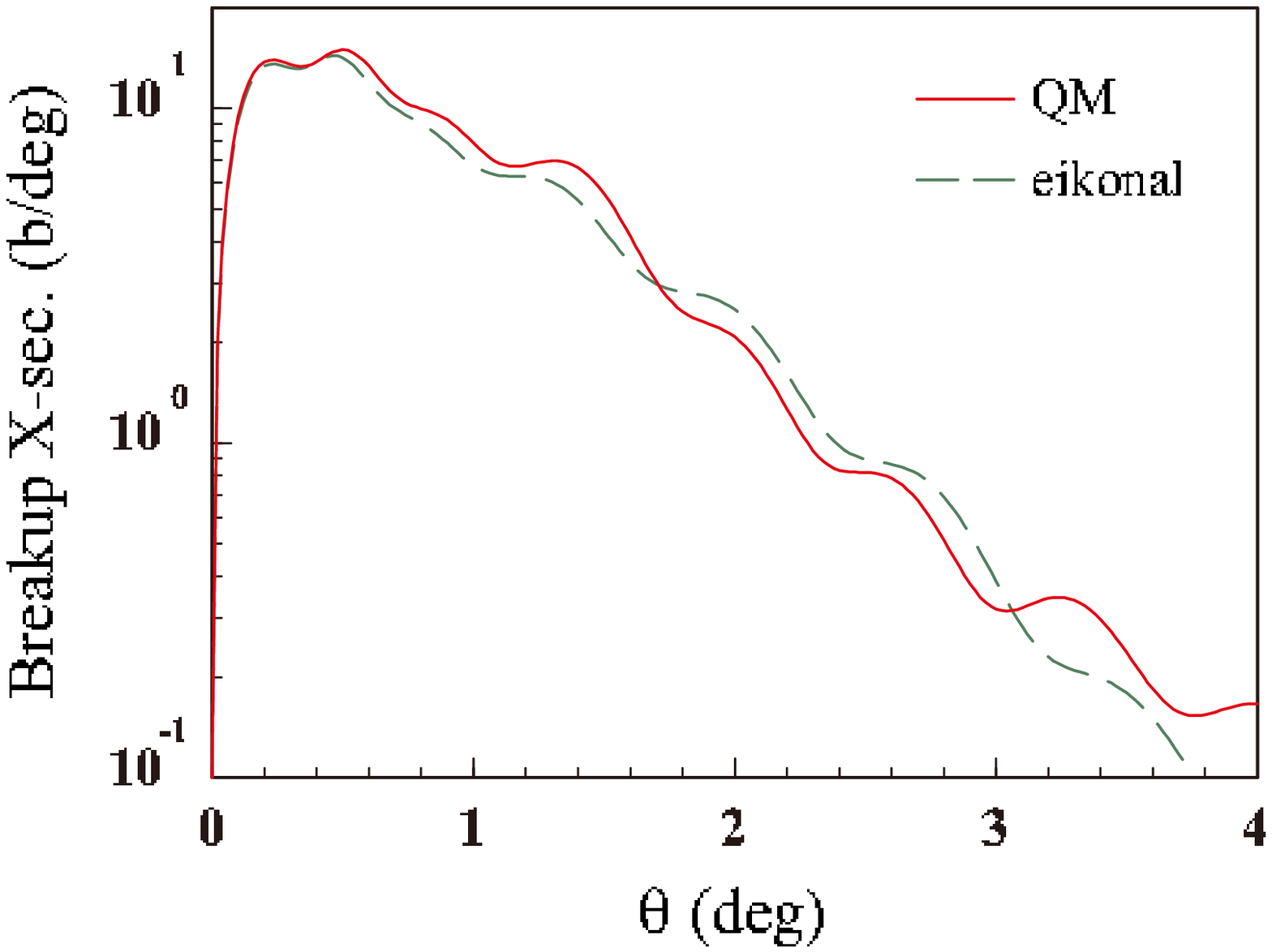}
\includegraphics[width=70mm,keepaspectratio]{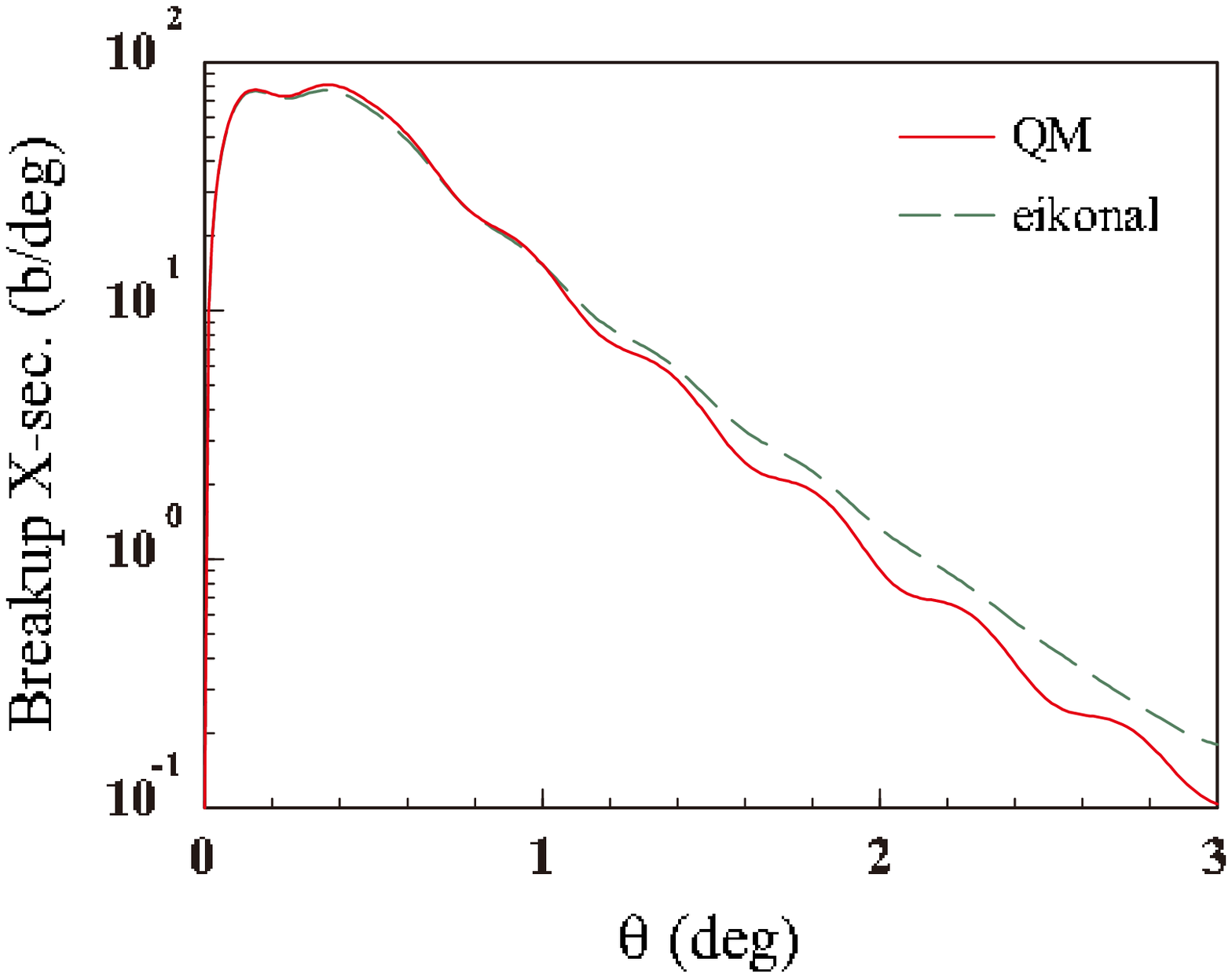}
} \caption{\label{fig12} Breakup cross sections for $^8$B$+^{208}$Pb
(left panel) and $^{11}$Be$+^{208}$Pb (right panel) at 250
MeV/nucleon as a function of the scattering angle. The solid and
dashed lines show the results of nonrelativistic quantum mechanical
CDCC and nonrelativistic eikonal CDCC, respectively.}
\end{figure}

We show in Fig.~10 the breakup amplitudes for $^8$B$+^{208}$Pb at
250 MeV/nucleon as a function of $L$. We choose the s-wave 6th bin
state, whose breakup amplitude has the largest value, as a final
state; the $z$-component $m_0$ of the spin of $^8$B in the incident
channel is chosen as 1. The upper and lower panels correspond to the
real and imaginary parts of the amplitude, respectively. The solid
line represents the result of nonrelativistic QM CDCC,
which adopts Eq.~\eqref{fCq} for all $L$ and
has no dynamical relativistic corrections.
The dotted and dashed lines are the
results of E-CDCC, based on Eq.~\eqref{fC7},
with and without dynamical relativistic corrections;
below we call the former relativistic E-CDCC and the latter
nonrelativistic E-CDCC.
One sees from the figure that at small
$L$, i.e., $L \la 500$, the dashed and dotted lines agree very well
with each other, and deviate from the solid line. On the other hand,
at large $L$, the solid and dashed lines show a very good agreement
and differ from the dotted line. This is indeed consistent with the
above mentioned properties of relativistic and QM corrections with
respect to $L$. Thus, using the amplitude obtained by
nonrelativistic QM CDCC for small $L$ and those by relativistic
E-CDCC for large $L$ allows one to construct an accurate CC
framework that includes dynamical relativistic corrections and
QM effects, i.e., relativistic CDCC adopted in
\S\ref{sec3-2} and \ref{sec3-3}.

The difference of the DDBUX with nonrelativistic QM CDCC
from that with nonrelativistic E-CDCC is
shown in Fig.~11. The left and right panels correspond to
$^8$B$+^{208}$Pb and $^{11}$Be$+^{208}$Pb at 250 MeV/nucleon,
respectively. Compared to the higher-order and nuclear breakup
corrections, the difference due to the QM effect, i.e., the {\it QM
correction}, seems very small. It is appreciable, however, in the
angular distribution of the BUX, as shown in Fig.~12. The solid and
dashed lines represent the BUX obtained with
nonrelativistic QM CDCC and nonrelativistic E-CDCC,
respectively. For both $^8$B$+^{208}$Pb (left panel) and
$^{11}$Be$+^{208}$Pb (right panel) reactions, the $\theta$
dependence of the BUX at angles above the peak position is indeed
sensitive to the QM correction.

\subsection{Role of close-field collisions}
\label{sec3-5}

Finally, we evaluate the validity of the far-field
approximation for
the Coulomb interaction used in this study. Figure 13 shows the
breakup amplitude multiplied by a weight value $b$. The left and
right panels show the results for $m_0=\mu=0$ and 1, respectively,
obtained with the first order perturbation model described in
\S\ref{sec2-2}. In each panel, the solid (dashed) line shows the
amplitude corresponding to both the close and far fields (only the
far field). It is clearly seen that the contribution from the close
field of the Coulomb interaction is negligibly small.
We expect that the inclusion of higher order processes and the
nuclear breakup does not change this conclusion.

%
\begin{figure}[b]
\centerline{
\includegraphics[width=70mm,keepaspectratio]{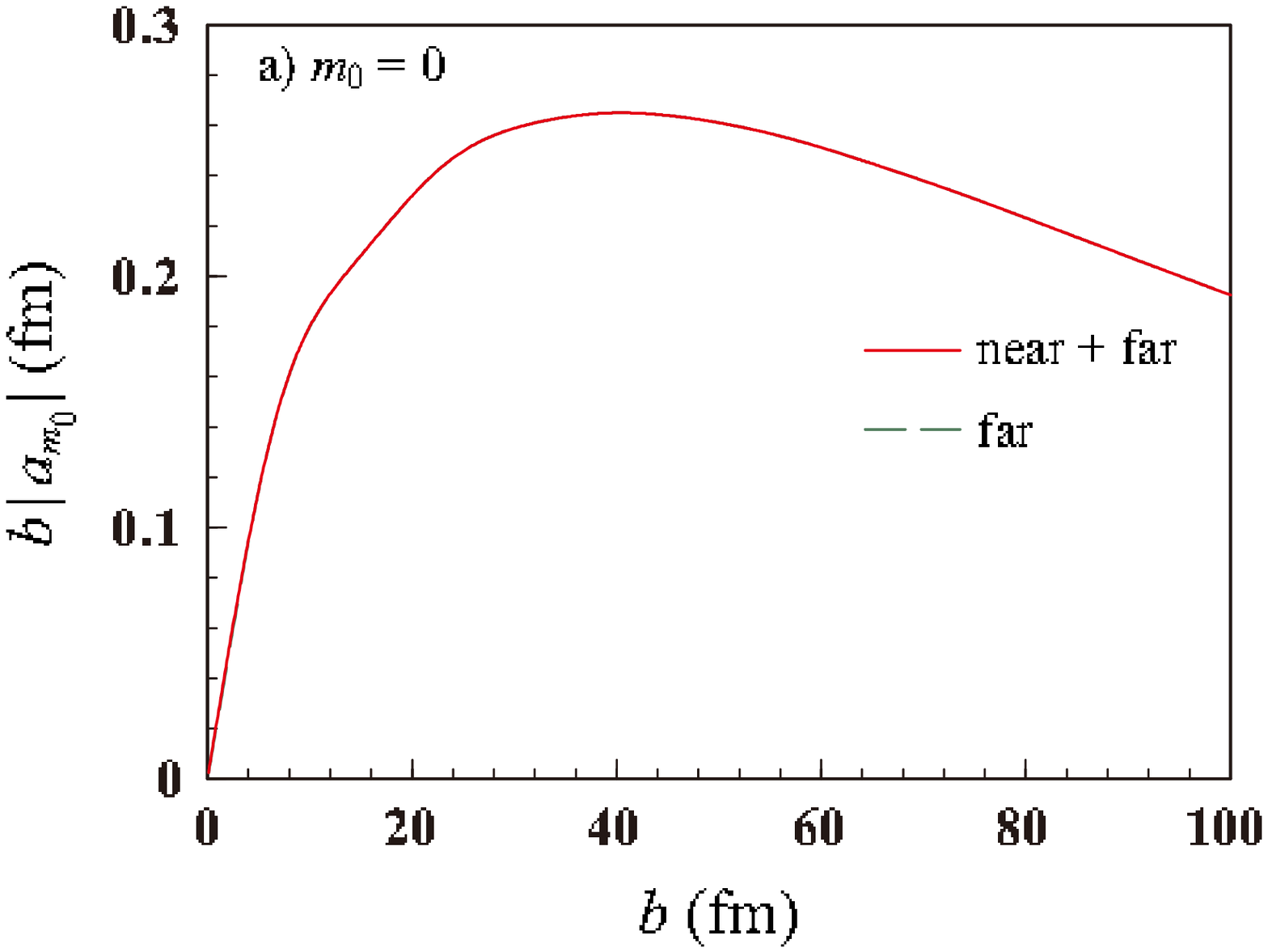}
\includegraphics[width=70mm,keepaspectratio]{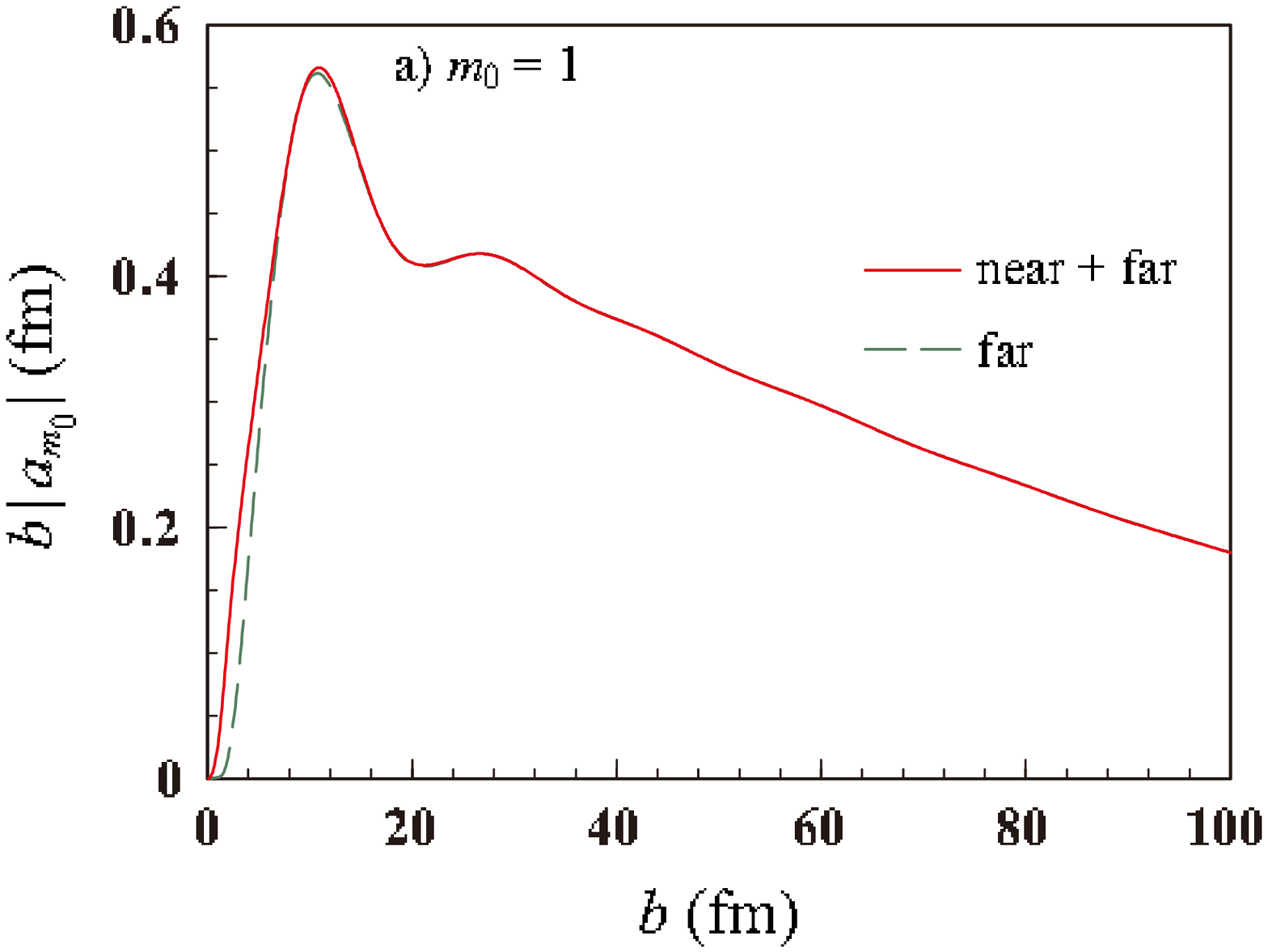}
} \caption{\label{fig13} Weighted breakup amplitude for
$^8$B$+^{208}$Pb at 250 MeV/nucleon as a function of the impact
parameter $b$. The final state of $^8$B is the same as in
Fig.~\ref{fig5}. The left and right panels correspond to $m_0=\mu=0$ and
1, respectively. In each panel, the solid (dashed) line shows the
result including both far and close contributions (only the far
contribution).}
\end{figure}
In Ref.~\citen{EB02-2} the close-field contribution
is shown to be important at lower energies, i.e., $E\la 50$ MeV/nucleon.
At these low energies, however, the dynamical relativistic
effects are very small and we do not need the far-field approximation
that is used only to obtain the transformation form of the
Coulomb coupling potentials due to relativity. Note that in the
evaluation of the nonrelativistic Coulomb potentials, Eq.~\eqref{NRC},
both close and far fields are taken into account.

\section{Summary}
\label{sec4}

The dynamical relativistic effects on the $^8$B and $^{11}$Be
double differential breakup cross sections (DDBUX)
at 250 MeV/nucleon are investigated in detail by means of
the Continuum-Discretized Coupled-Channels method (CDCC).
The effects on the DDBUX are indeed large, i.e., several tens
of \% at forward angles $\theta$ and at slightly higher breakup
energies $\epsilon$ than at the peak of the breakup energy spectrum.
The contribution of the nuclear breakup and higher-order processes
are found to be the order of 10\% and has different $\theta$-$\epsilon$
dependence between the $^8$B and $^{11}$Be breakup processes.
It is confirmed that the relativistic corrections are only
appreciable for breakup amplitudes corresponding to large
impact parameters, and quantum-mechanical correction is
negligible there. This feature enables one to perform
fully relativistic and quantum mechanical coupled-channel
calculations of breakup reactions at intermediate energies.
The far-field approximation, which is used in the
formulation of relativistic Coulomb coupling potentials,
is justified numerically by means of a first-order perturbative
calculation. Thus, we now have an accurate method to analyze
experimental data of nuclear and Coulomb breakup reactions at
intermediate energies by means of {\it relativistic CDCC}.
Inclusion of magnetic transitions will be an important future work.

\section*{Acknowledgements}
This work was partially supported by the U.S. DOE grants DE-FG02-08ER41533 and
DE-FC02-07ER41457 (UNEDF, SciDAC-2), the Research Corporation,
and the JUSTIPEN/DOE Grant DEFG02-06ER41407.
The computation was carried out using the computer facilities at
the Research Institute for Information Technology, Kyushu University.


\end{document}